\documentclass[runningheads]{svmult}

\usepackage{makeidx}   % allows index generation
\usepackage{graphicx}  % standard LaTeX graphics tool
\usepackage{subeqnar}  % subnumbers individual equations
\usepackage{multicol}  % used for the two-column index
\usepackage{physprbb}  % modified textarea for proceedings,
\makeindex             % used for the subject index
\usepackage{theorem}
\usepackage{amssymb}
\theoremstyle{break} \newtheorem{Def}{Definition}

\DeclareFontFamily{OT1}{rsfs}{}
\DeclareFontShape{OT1}{rsfs}{m}{n}{ <-7> rsfs5 <7-10> rsfs7 <10->
rsfs10}{} \DeclareMathAlphabet{\mycal}{OT1}{rsfs}{m}{n}
\begin{document}
\title*{Numerical relativity with the conformal field equations}
\toctitle{Numerical relativity with the
\protect\newline  conformal field equations}
% allows explicit linebreak for the table of content
%
%
\titlerunning{Numerical relativity with the conformal field equations}
% allows abbreviation of title, if the full title is too long
% to fit in the running head
%
\author{Sascha Husa\inst{1}}
\authorrunning{Sascha Husa}
% if there are more than two authors,
% please abbreviate author list for running head
%
%
\institute{Max-Planck-Institut f\"ur Gravitationsphysik, 14476 Golm, Germany}

\maketitle              % typesets the title of the contribution

\begin{abstract}
I discuss the conformal approach to the numerical
simulation of radiating isolated systems in general
relativity. The method is based on conformal compactification and
a reformulation of the Einstein equations in terms of rescaled variables,
the so-called ``conformal field equations'' developed by Friedrich.
These equations allow to include ``infinity'' on a finite grid, solving
regular equations, whose solutions give rise to solutions of the Einstein
equations of (vacuum) general
relativity. The conformal approach promises certain advantages, in particular
with respect to the treatment of radiation extraction and boundary conditions.
I will discuss the essential features of the analytical approach to the
problem, previous work on the problem -- in particular
a code for simulations in 3+1 dimensions, some new results,
open problems and strategies for future work.
\end{abstract}

\section{Introduction}\label{sec:intro}
%%%%%%%%%%%%%%%%%%%%%%

In order to understand the physical content of the theory of general
relativity, it is desirable
% to understand its solutions
% both {\em mathematically} and
%{\em observationally}, and both pursuits are indeed tightly coupled.
to both {\em mathematically} understand its solutions and
{\em observationally} understand the physical phenomena for which
the theory is relevant.
The latter effort typically requires predictions from the theory,
both qualitative and quantitative -- such
as gravitational wave templates or binary pulsar deceleration parameters.
The lack of genericity in available exact solutions then naturally leads to
the use of approximation methods such as post-Newtonian approximations,
perturbation theory or numerical analysis, which allows very general
non-perturbative approximations.
Concrete solutions do however also play an important role in the quest for
a mathematical understanding of the solution space. The experience
gained from such solutions can suggest theorems, test conjectures, or lead to
the discovery of previously unknown phenomena.
For some particularly interesting examples see
\cite{choptuik}, \cite{3d_solitons} or \cite{marsha+alan}.
The construction and study of solutions, be it with approximate or
exact methods, obviously profits from a sound mathematical basis in the
form of well-posed equations, analytic estimates and the likes. Eventually 
-- hopefully -- it will also profit from observational evidence!

In the following I will discuss a particular approach to the numerical
solution of the Einstein field equations, which addresses the problems
associated with the treatment of asymptotic regions by conformal
compactification.
The interest in asymptotic regions is rooted in the problem of
describing isolated systems. Physical intuition suggests that 
many astrophysical processes (whether they are of actual astrophysical
relevance or rather hypothetical)
should essentially be independent of the large-scale structure of the
universe, or, say, the local galaxy.
The idealization of an isolated system, where
the geometry approaches a Minkowski geometry at large
distances, thus forms the basis for the general-relativistic analysis of 
processes which are essentially of non-cosmological nature.
The mathematical formalization of
the physical idea of isolated systems is the concept of
{\em asymptotically flat spacetimes}.
This formalization is already nontrivial, due to the lack of a preferred
background geometry or coordinate system -- with respect to which one could
define ``distance'' and the appropriate limits.
Conformal compactification, however,
renders possible a discussion of asymptotically
flat spacetimes in terms of local differential geometry.
In this approach, pioneered by Penrose \cite{Pe63ap}, an
unphysical Lorentzian metric $g_{ab}$ is introduced on an unphysical manifold
${\cal M}$ which gives rise to the physical metric $\tilde g_{ab}$
by the rescaling $\tilde g_{ab} = \Omega^{-2} g_{ab}$. The physical
manifold $\tilde {\cal M}$ is then given by $\tilde {\cal M} =
\{ p \in {\cal M} \, \vert \, \Omega(p) > 0  \}$.
In this picture physical ``infinity'' corresponds to a three-dimensional
boundary of a four-dimensional region in ${\cal M}$, defined by $\Omega=0$.
Limiting procedures and approximations
can thus be replaced by local differential geometry on the boundary.

In gravitational theory, quantities such as the total mass, (angular)
momentum or emitted gravitational radiation can only consistently be defined
at ``infinity''.
In the conformal approach the unambiguous extraction of gravitational waves
from a numerical spacetime is straightforward. In the ``traditional''
approach to dealing with asymptotic falloff in numerical relativity,
where one introduces an arbitrary spatial cutoff, matters are much more
complicated and ambiguities are introduced which one would have to get rid off
by complicated limiting procedures. 
Without at least being able to define a clean concept of radiation leaving or 
entering a system, it is furthermore very hard to define physically realistic
and consistent boundary conditions at finite distance.
%By choosing the
%the grid boundary to be located at our beyond  $\Omega=0$
%the physical region gets causally (if not numerically) disconnected from the
%grid boundary, and such ambiguities are removed (or more precisely,
%the associated errors should converge to zero consistently with the error
%of the discretization scheme).
The traditional approach is thus not completely satisfactory both
from a mathematical but also from a practical 
point of view. Here we discuss the principal ideas of the envisioned
``conformal cure'', the technical and conceptual problems
associated with it, and the current status of this approach.

It is easy to see that the conformal cure can not be straightforward,
by writing Einstein's vacuum equations in terms of $\Omega$ and $g_{ab}$:
\begin{equation}\label{eq:conformalG_naive}
\tilde G_{ab}[\Omega^{-2} g_{ab}] =  G_{ab}[g_{ab}] + \frac{2}{\Omega}
\left(\nabla_a \nabla_b\Omega + g_{ab} \nabla_c \nabla^c \Omega \right)
 +\frac{3}{\Omega^2} g_{ab}\left(\nabla_c \Omega \right)\nabla^c \Omega\; .
\end{equation}
This expression is singular for $\Omega=0$, multiplication by $\Omega^2$
also does not help here because then the principal part of the
partial differential
equations encoded in $G_{ab}$ would degenerate at  $\Omega=0$.
The conformal compactification approach thus can not be
carried to the level of the field equations in a straightforward way.
This step however has been
achieved by Friedrich, who has developed a judicious reformulation of the
equations 
\cite{CFE_ProcRSoc81I,CFE_ProcRSoc81II,CFE_Commun83,CFE_JGeometry98,CFE_GaussianGauge}. 
These {\em conformal field equations} are {\em regular}
equations for $g_{ab}$ and certain additional independent variables.

In analytical work, such global methods have proven to provide
essential simplifications leading to new results and insights. 
Already by providing a different point of view on some of the essential
problems in numerical relativity, the conformal picture is quite helpful
and can stimulate new ideas. Certainly, we desire more -- to make
this approach also a practical tool. There is significant hope, that
global methods will eventually show advantages for practical
numerical work, and despite the small number of researchers involved so far
(may there be more!), some significant progress in this direction has been
made.

In the present article I will try to sketch the present status of the quest
for the conformal cure and discuss some important open questions.
We will start with a brief introduction of the concepts of asymptotic
flatness in terms of conformal compactification in Sec.
\ref{sec:isolated_systems}, highlighting some important features
of ``future null infinity'', and then discuss the conformal field equations.
In Sec. \ref{sec:examples} I will discuss some explicit examples of
compactifying Minkowski spacetime, both to paint a more concrete picture of
our scenario, and to set the arena for some numerical code tests.
Sec. \ref{sec:history} contains a brief overview of the history of
numerical work on the conformal field equations, leading to a description
of a 3D code written by H\"ubner \cite{PeterI,PeterII,PeterIII,PeterIV}.
New results from 3D calculations performed with this code will be presented in
Sec. \ref{sec:results}, and a discussion will be given in Sec.
\ref{sec:discussion}, concluding with a roadmap for future work.

\section{Compactification and the Mathematical Description of Isolated Systems}
%%%%%%%%%%%%%%%%%%%%%%%%%%%%%%%%%%%%%%%%%%%%%%%%%%%%%%%%%%%%%%%%%%%%%%%%%%%%%%%
\label{sec:isolated_systems}

The material in this section is intended to present some essential ideas in
a condensed form. The reader should be aware that I am not
doing justice here to subtleties and long history of the mathematical
description of isolated systems in general relativity --
rather this section intends to motivate to look into more
complete reviews such as \cite{Tod+Newman,Wald,Stewart}.

\subsection{Asymptotic Flatness and Compactification}
\label{sec:compactification}
%%%%%%%%%%%%%%%%%%%%%%%%%%%%%%%%%%%%%%%%%%%%%%%%%%%%%%%%%%%%%%%%%%%%%%%%

As noted above the formulation of the concept of asymptotic flatness is far
from straightforward in GR, due to the absence of a background metric or
preferred coordinate system, in terms of which falloff rates can be
specified.
A resolution of this problem is provided by a definition of asymptotic
flatness, where, after a suitable conformal rescaling of the metric, ``points
at infinity'' are added to the manifold. One thus works on a compactified
auxiliary manifold, and local differential geometry can be used to study
the asymptotic properties of the gravitational field.
We will give a simple definition of asymptotic flatness here, which for
our purposes catches all essential features. For alternative definitions
and more detailed explanations compare for example
\cite{Pe63ap,Wald,Tod+Newman,joerg_livrev}.

\begin{Def}[asymptotic simplicity]\label{def:as.simple}
A smooth spacetime $(\tilde{\cal M}, \tilde g_{ab})$ is
called asymptotically simple, if there exist another smooth spacetime
$({\cal M}, g_{ab})$ and a scalar function $\Omega$ such that:
\begin{enumerate}
\item $\tilde{\cal M}$ is an open submanifold of ${\cal M}$ with
smooth boundary $\partial\tilde{\cal M} = {\mycal I}$ (Scri).\\
\item  $g_{ab} = \Omega^2 \tilde g_{ab}$ on $\tilde{\cal M}$, with
$\Omega > 0$ on  $\tilde{\cal M}$, $\Omega=0$ on ${\mycal I}$ and
  $\nabla_a \Omega\neq0$ on ${\mycal I}$.\\
\item 
Every null geodesic in $\tilde {\cal M}$ acquires two end points on ${\mycal I}$.
\end{enumerate}
\end{Def}
\begin{Def}[asymptotic flatness]\label{def:as.flat}
Asymptotically simple spacetimes %$(\tilde{\cal M}, \tilde g_{ab})$ 
are called asymptotically flat
if their Ricci tensor $\tilde R_{ab}$ vanishes in a neighborhood of ${\mycal I}$.
\end{Def}  

Examples of asymptotically simple spacetimes, which are not asymptotically
flat are the de Sitter and anti-de Sitter solutions. Correspondingly to
asymptotically flat spacetimes one can consider asymptotically
de Sitter and anti-de Sitter spacetimes.
Note that the completeness condition 3 in Def. \ref{def:as.simple},
which ensures that the entire boundary is included, excludes
black-hole spacetimes. For modifications to weaken condition 3,
thus allowing black holes, 
see the definitions of \cite{Penrose} or \cite{Wald}.
For example, the definition of {\em weak asymptotic simplicity}
\cite{Penrose} requires condition 3 to hold only in a neighborhood
of ${\mycal I}$. See e.g.  \cite{Wald} for a discussion of asymptotic flatness
at spacelike infinity (i.e. the part of infinity which is reached
along spacelike geodesics)
versus null infinity (i.e. the part of infinity which
is reached along null curves).
The notion of asymptotic flatness at timelike infinity does not make
much sense in a general situation, because then all energy would have to be
radiated away, leaving only flat space behind -- excluding black holes or
``stars''. For weak data however, in vacuum say, where all radiation eventually
disperses, once expects asymptotic flatness to hold also at timelike infinity,
this issue will be discussed below in application to concrete spacetimes.

The notion of asymptotic flatness of isolated systems turns out to be
intimately related to the possibility of defining the total energy-momentum
for such systems in general relativity --
remember that no well-defined local energy density of the
gravitational field is known (compare e.g. Sec. 11.2 of the
textbook of Wald \cite{Wald}). However, total energy-momentum quantities,
which transform as a 4-vector under asymptotic Lorentz transformations, can be
assigned to null and spatial infinity of asymptotically flat spacetimes.
If a manifold has more than one asymptotically flat end, e.g. in the
presence of wormholes of the Einstein-Rosen-bridge type, then
different energy-momenta can be associated with each of these asymptotic
regions.

The expression for the energy momentum four-vector at spatial infinity
has been given first by Arnowitt, Deser and Misner in 1962 \cite{ADM}
in the context of the Hamiltonian formalism, and is usually called the ADM
momentum, the time component being called ADM mass.
The ADM energy corresponds to the energy of some Cauchy surface,
i.e. a snapshot of the spacetime at some fixed time.
It is a constant of motion and can therefore be expressed in terms of the
initial data on an asymptotically flat Cauchy hypersurface.

The expression for the energy momentum at null infinity,
usually referred to as the Bondi energy-momentum, can be associated
with a fixed retarded time, i.e. some asymptotically null surface.
The decrease of this quantity measures the energy-momentum carried away by
gravitational radiation.
For a brief introduction and references to original work on different
definitions of the Bondi mass see e.g. the textbook of Wald \cite{Wald}. 
The formulation most appropriate for usage in numerical codes based on the
conformal field equations was given by Penrose \cite{Pe63ap},
and defines the Bondi mass in terms of the behavior of certain projections
of the Weyl tensor at ${\mycal I}^+$ and the shear of the outgoing congruence
of null geodesics orthogonal to ${\mycal I}$ in the gauge defined below by 
(\ref{eq:Bondi_conformal_gauge}).
It was already shown in 1962 by Bondi, van der Burg and Metzner
\cite{Bondi_et_al} that the Bondi mass $M_B$ can
only {\em decrease} with time: gravitational radiation always
carries positive energy away from a radiating system.
Note that this means in particular, that
while compactification at spatial infinity would lead to a ``piling up'' of 
waves, at ${\mycal I}^+$ this effect does not appear. In the compactified
picture the waves leave the physical spacetime through the boundary
${\mycal I}^+$.

A fundamental issue of general relativity
is the positivity of the ADM and Bondi energies.
Although it is trivial to write down a metric with negative mass
if no conditions on the energy momentum tensor are imposed,
for {\em reasonable} matter fields with nonnegative energy density (thus
satisfying the dominant energy condition), non-negativity of the ADM and Bondi
energies is expected on physical grounds:
if the energy of an isolated system could be negative, it would most
likely be unstable and decay to lower and lower energies.
Indeed, a proof of the positive definiteness of the ADM energy
has been given in 1979 by Schoen and Yau \cite{sy} (several simplified proofs
have been given later), and was extended to the Bondi mass in 1982 by
Horowitz and Perry \cite{Horowitz_Perry}.

\subsection{What is Scri?}\label{sec:scri}
%%%%%%%%%%%%%%%%%%%%%%%%%%%%%%%%%%%%%%%%%%%%%%%%%%%%%%%%%%%%%%%%%%%%%%%%

We will now have a closer look at ${\mycal I}$ and discuss some of its features,
which will allow us to understand the basic ideas of radiation extraction and
help us to understand some issues related with choosing boundary conditions
for numerical solutions of the conformal field equations.

Looking at  (\ref{eq:conformalG_naive}) and multiplying by $\Omega^2$,
one can see that for a vacuum spacetime, $\tilde G_{ab} = 0$,
$(\nabla_c \Omega) \nabla^c \Omega = 0$ at 
${\mycal I}$, which thus must consist of null surfaces.
In fact, one can then prove (see e.g. \cite{joerg_livrev}), that
\begin{enumerate}
\item ${\mycal I}$ has two connected components, each with topology $S^2 \times R$.\\
\item The connected components of ${\mycal I}$ are smooth null hypersurfaces
 in ${\cal M}$, and as such are generated by null geodesics.\\
\item The congruence of null geodesic generators of ${\mycal I}$ is shear free.
\end{enumerate}
The two connected components are called future null infinity (${\mycal I}^+$)
and past null infinity (${\mycal I}^-$), and provide the future and
past endpoints for null geodesics in $\tilde {\cal M}$. In a naive picture
they could be viewed as emanating from a point $i^0$ which represents
spatial infinity \footnote{The structure of $i^0$ is however quite subtle,
significant progress toward its understanding in terms of the field
equations has recently been achieved by Friedrich 
\cite{CFE_JGeometry98,CFE_Pune98}}.
These features will become more graphic when dealing with
explicit examples below.

Note that there is gauge freedom in the choice of the conformal
factor: one is free to rescale the conformal factor $\Omega$ by some
$\omega > 0$ such that $\hat\Omega = \omega \Omega$,
$\hat g_{ab} = \omega^2 g_{ab} = \hat\Omega^2 \tilde g_{ab}$.
It is an interesting exercise (see Sec. 11.1 of  \cite{Wald}) to prove
that outside any neighborhood of $i^0$ -- on ${\mycal I}^+$ say --
one can always use this conformal gauge freedom to achieve
\begin{equation}\label{eq:Bondi_conformal_gauge}
\hat\nabla_a \hat\nabla_b \hat\Omega = 0 \qquad \mbox{on ${\mycal I}^+$} \; ,
\end{equation}
where $\hat\nabla_a$ is the derivative operator compatible with the metric
$\hat g_{ab}$.
This conformal gauge implies, that the null tangent $n^a = \hat g^{ab}\nabla_b \hat \Omega$ to the null geodesic generators of
${\mycal I}$ satisfies the affinely parameterized geodesic equation,
\begin{equation}\label{eq:geodesic_generators}
n^a \hat\nabla_a \hat n^b = 0 \; .
\end{equation}
Consequently, expansion of the generators of ${\mycal I}$ vanishes in
addition to the shear and twist ($n_a$ is a gradient).
Using the remaining gauge freedom of $\omega$, we can choose
coordinates such that the metric on ${\mycal I}$ takes the form
\begin{equation}\label{eq:scri_metric}
  \D {\hat s}^2 \vert_{{\mycal I}^+} =
          2\D \Omega \, \D u + \D \theta^2 + \sin^2\theta \D \phi^2 \; ,
\end{equation}
where $u$ is the affine parameter of the null geodesic generators, scaled
such that $n^a \hat\nabla_a u = 1$ (see e.g. Chpt. 11 of \cite{Wald}).
The cuts \footnote{A cut of ${\mycal I}$ is a two-dimensional spacelike
cross section of ${\mycal I}$ which meets every generator once.} of ${\mycal I}$
of constant $u$ thus become metric spheres. The coordinate $u$ is
generally known as Bondi parameter or Bondi time.
The conformal gauge (\ref{eq:Bondi_conformal_gauge}) and the coordinates
 (\ref{eq:scri_metric}) prove
very useful in the analysis of the geometry in a neighborhood of
${\mycal I}$ --  in particular for the extraction of radiation.
The existence of a natural time coordinate (at least up to affine
transformations along {\em each} generator) is very interesting
for numerical applications, where at least asymptotically one
can get rid of much of the slicing arbitrariness of the interior 
region. It is nontrivial but rather straightforward to actually
(numerically) find this gauge of ${\mycal I}^+$, which is also required
by the standard formulas to compute the
energy-momentum at ${\mycal I}^+$ and the emitted radiation -- to
be given below.

Before discussing how to compute the radiation, it is useful to
idealize a detector (here I will follow the discussion in \cite{JoergIV}).
In physical space -- far away form the sources --
we could think of a detector as a triad of spacelike unit vectors attached
to the worldline of some (timelike) observer. Let us further assume
for simplicity that the observer moves along a timelike geodesic parametrized
by proper time and that the triad is transported by Fermi-Walker transport.
It is not hard to show -- see Frauendiener \cite{JoergIV}, that taking
the appropriate limit in the compactified spacetime, the observer worldline
converges to a null geodesic generator of ${\mycal I}^+$. Taking the limit
along a Cauchy surface it converges to the point $i^0$,
where one could naively expect
an observer to end up when shifted to larger and larger distances
(this limit is however not appropriate in the context of computing
the radiation).
Furthermore, the proper time parameter of the observer converges to
Bondi time. The arbitrariness of boosting the observers is reflected in
the affine freedom of choosing the Bondi parameter at ${\mycal I}$.
The description of ${\mycal I}^+$ thus could be condensed into the statement
that it idealizes {\em us} -- the observers of astrophysical phenomena
happening far away. By working with the idealization, the approximations
and ambiguities associated with detectors at a finite distance have
transformed into a surprisingly simple geometric picture!
Note that this simplification has to be taken with the typical
care required in the treatment of idealizations in (theoretical)
physics: Under practical circumstances, e.g. computing the actual
signal at a gravitational wave detector, ${\mycal I}$ more realistically
corresponds to an observer that is sufficiently far way from the source
to treat the radiation linearly, but not so far away that cosmological
effects have to be taken into account. In order to compute the detected
signal in a realistic application,
cosmological data and the fact that an earthbound detector
moves in a complicated way relative to the source all have to be considered.

We will next discuss a ``detector-frame'' adapted to ${\mycal I}^+$ -- the
commonly used Bondi frame. For a much more complete discussion of Bondi-systems
see e.g. the excellent review by Newman and Tod \cite{Tod+Newman}.
There a characteristic framework is used to set up the Bondi frame in
a whole neighborhood of ${\mycal I}^+$, which is necessary to compute derivatives,
entering e.g. the definition of the spin coefficient $\sigma$ defined
below in  (\ref{eq:def_sigma}). In the current approach, the Bondi system
is only defined at ${\mycal I}^+$: initial data can be set up, such that
all necessary quantities can be propagated along the generators of
${\mycal I}^+$ \cite{unpublished}.

With  ${\mycal I}^+$ being a null surface, it is most
natural to use a null frame, consisting of 2 null vectors and 2 spacelike
vectors $x^{a}$, $y^{a}$, which can be considered as the idealizations
of the arms of an interferometric gravitational wave detector.
The vectors  $x^{a}$, $y^{a}$ are
commonly treated in the form of two complex null vectors $m^a, \bar m^a$, with
$$
m^a = x^a + i y^a, \qquad m^a \bar m_a = 1 \; ,
$$
where $x^a$ and $y^a$ are real vectors tangent to the cuts of ${\mycal I}^+$.
The null vectors are taken as the affine tangent $n^a$ and
$l_a = \hat\nabla_a u$, which satisfy
$$
n^a l_a = -1 \; .
$$
The tetrad vectors $l_a$, $m^a$ and $\bar m^a$ are parallely propagated along
the generators, which yields transport equations that define them on
all of ${\mycal I}^+$ once initial values are chosen.

The Bondi-mass can then be computed in terms of the spin-coefficient $\sigma$
and the rescaled Weyl tensor components $\psi_2$ and $\psi_4$:
\begin{eqnarray}
\sigma &=& \hat g_{ab} l^a m^c \hat\nabla_c m^b    \; , \label{eq:def_sigma} \\
\psi_2 &=& \hat d_{abcd} l^a m^b \bar m^c n^d      \; , \label{eq:def_psi2}  \\
\psi_4 &=& \hat d_{abcd} n^a \bar m^b n^c \bar m^d \; .\label{eq:def_psi4}
\end{eqnarray}
In terms of these quantities the Bondi mass can be defined as
\begin{equation}
M_B = -\frac{ \sqrt{A}}{\sqrt{4\pi}^3}
 \int\left(  \psi_2 + \sigma\dot{\bar\sigma}\right) \, dA \; ,
\end{equation}
the outgoing radiation can be computed to be
\begin{equation}
\dot M_B = -\frac{ \sqrt{A}}{\sqrt{4\pi}^3}
 \int\left(  \dot\sigma\dot{\bar\sigma}\right) \, dA \; ,
\end{equation}
where $A$ is the area of the cuts of ${\mycal I}^+$, and $\dot f = n^a \hat\nabla_a
 f = \partial_u f$.
Furthermore, 
$$
\ddot \sigma = - \bar\psi_4
$$
can be used to evolve $\sigma$, where both $\sigma$ and $\dot\sigma$
can be computed on the initial slice.

This procedure has been implemented by H\"ubner and  Weaver \cite{unpublished}
for 2D codes and the 3D code used to obtain the results in Sec.
\ref{sec:results}, and has been tested and proven accurate for
several types of spacetimes \cite{unpublished}.
Frauendiener describes his implementation and some results in \cite{JoergII}.
There are two essential problems in these implementations: First of all,
the gauge conditions will not usually result in a slicing of ${\mycal I}^+$
by cuts of constant Bondi time $u$. This means that interpolation has to be
used between different slices of the numerical evolution. Second, 
in those formulations of the conformal field equations that
have so far been used in numerical implementations, the conformal factor
$\Omega$ is an evolution variable and not specified a priori,
${\mycal I}$ will in general not be aligned with grid points. This results
in further technical complications and an additional need for interpolation.
When dealing with the physically interesting case of a ${\mycal I}^+$ of
spherical topology, at least two patches have to be used to represent
the Bondi tetrad $(l_a,n_a,m_a,\bar m_a)$.
Frauendiener has achieved to control the movement of ${\mycal I}^+$
through the grid by the gauge choice  for his formulation
\cite{JoergII}, in particular the shift vector can be chosen such that
${\mycal I}$ does not change its coordinate location.

\subsection{The Conformal Field Equations}\label{sec:CFE}
%%%%%%%%%%%%%%%%%%%%%%%%%%%%%%%%%%%%%%%%%%%%%%%%%%%%%%%%%%%%%%%%%%%%%%%%

Several formulations of the conformal field
equations are available, the main difference being whether
the conformal factor $\Omega$ can be specified a priori or
is determined as a variable by the equations.
In the original formulation \cite{CFE_ProcRSoc81I,CFE_ProcRSoc81II}
and its descendants \cite{CFE_Commun83,PeterI,JoergI} $\Omega$ (and
derivatives) are evolved as dependent variables. All existent numerical codes
are based on equations of this type. 
A later version of the equations allows to fix $\Omega$ a priori, and has been
used to develop a new treatment of spatial infinity $i^0$
\cite{CFE_JGeometry98,CFE_Pune98,CFE_GaussianGauge}.
However, the formulation and treatment of these equations is more involved,
and its numerical solution has not yet been attempted.

In the following we discuss a metric based formulation of the
``original'' version of the conformal field equations, which forms
the basis for H\"ubner's codes \cite{PeterI,PeterII,PeterIII,PeterIV}.

When deriving the conformal field equations, it turns out to be useful to
start with the splitting of the Riemann tensor into its trace-free (the Weyl
tensor) and trace (Ricci tensor and scalar) parts. Additionally
we define the tracefree Ricci tensor
$\hat R_{ab} = R_{ab} - \frac{1}{4}\, g_{ab}\, R$
and the rescaled Weyl tensor
\begin{equation}
d_{abc}{}^d =  \Omega^{-1} \, C_{abc}{}^d \; .
\end{equation}

The requirement that the physical scalar curvature $\tilde R$ vanishes implies
\begin{equation}\label{eq:Rtrace}
  6 \, \Omega \, \nabla^a \nabla_a \Omega =
    12 \, (\nabla^a \Omega) \, (\nabla_a \Omega) - \Omega^2 R \; ,
\end{equation}
Note that this equation is {\em not} manifestly regular at $\Omega=0$, but 
it is actually possible to show that if  (\ref{eq:Rtrace}) is  satisfied
at one point, then by virtue of the other equations 
(\ref{eq:OmEq},\ref{eq:omWaveEq},\ref{eq:irrRiemann},\ref{eq:div_d},\ref{eq:TFRicci})
to be given below, it has to be satisfied everywhere. The whole system
(\ref{eq:Rtrace},\ref{eq:OmEq},\ref{eq:omWaveEq},\ref{eq:irrRiemann},
\ref{eq:div_d},\ref{eq:TFRicci}) is then regular in the sense that
this point does not have to be located at ${\mycal I}^+$.
The vacuum Einstein equations $\tilde R_{ab} = 0$ then yield
\begin{equation}\label{eq:OmEq}
\nabla_a \nabla_b \Omega 
 = \frac{1}{4} g_{ab} \, \nabla^c\nabla_c\Omega - \frac{1}{2}
\, \hat R_{ab} \, \Omega \; .
\end{equation}
Finally, commuting covariant derivatives in the expression
$$
g^{bc} \nabla_c \nabla_b \nabla_a \Omega
$$
and then using  (\ref{eq:OmEq}) again yields
\begin{equation}
\label{eq:omWaveEq}
\frac{1}{4} \nabla_a \left(\nabla^b\nabla_b \Omega\right) = -
\frac{1}{2} \, \hat R_{ab} \, \nabla^b \Omega 
- \frac{1}{24} \, \Omega \, \nabla_a R - \frac{1}{12} \, \nabla_a
\Omega \, R \; .
\end{equation}

Equations for the metric can be obtained by the identity
\begin{eqnarray} \label{eq:irrRiemann}
  R_{abc}{}^d &=&  \Omega d_{abc}{}^d
  + \left( g_{ca} \hat R_b{}^d - g_{cb} \hat R_a{}^d 
  - g^d{}_a \hat R_{bc} + g^d{}_b \hat R_{ac} \right)/2 \nonumber \\
& &   + \left( g_{ca} g_b{}^d - g_{cb} g_a{}^d \right) \frac{R}{12} \; ,
\end{eqnarray}
which defines the Weyl tensor. Expressing the
Riemann tensor  $R_{abc}{}^d$ in terms of the metric
and its derivatives (or the Christoffel quantities in a first
order formalism) yields the desired equations.
Note that for the physical Riemann tensor the
vacuum Einstein equations imply $\tilde  R_{abc}{}^d = \tilde C_{abc}{}^d$.

We still miss differential equations for $d_{abc}{}^d$ and $\hat R_{ab}$.
These can be obtained from the Bianchi identities
$\nabla_{[a} R_{bc]d}{}^e$, which
in terms of the Weyl and tracefree Ricci tensors imply
\begin{equation}\label{eq:Bianchi_Weyl}
\nabla_d \tilde C_{abc}{}^d = 0
\end{equation}
for the Weyl tensor of a vacuum spacetime ($\tilde R_{ab}=0$) and
\begin{equation}\label{eq:div_hatR}
\nabla_b \hat R_{a}{}^{b} = \frac{1}{4} \, \nabla_a R \; .
\end{equation}
While the Weyl tensor is conformally invariant,
$$
\tilde C_{abc}{}^d = C_{abc}{}^d \; , 
$$
this invariance does not hold for  (\ref{eq:Bianchi_Weyl}), instead
however one can show that
$$
\tilde\nabla_d \tilde C_{abc}{}^d =
                      \Omega \nabla_d \left(d_{abc}{}^d\right) \; , 
$$
which implies
\begin{equation}\label{eq:div_d}
\nabla_e d_{abc}{}^e = 0 
\end{equation}
if the vacuum Einstein equations hold in the physical spacetime.

The Bianchi identity combined with the splitting (\ref{eq:irrRiemann}) implies
\begin{equation}
\label{eq:TFRicci}
\nabla_a \hat R_{bc} - \nabla_b \hat R_{ac} = -
 \frac{1}{12} \left( (\nabla_a R) \, g_{bc} - (\nabla_b R) \, g_{ac} \right)
 - 2 \, (\nabla_d \Omega) \, d_{abc}{}^d \; .
\end{equation}

The equations (\ref{eq:Rtrace},\ref{eq:OmEq},\ref{eq:omWaveEq},\ref{eq:irrRiemann},\ref{eq:div_d},\ref{eq:TFRicci})
then constitute the conformal field equations for vacuum general relativity. 
Here the Ricci scalar $R$ of $g_{ab}$ is considered a given function
of the coordinates.
For any solution $(g_{ab},\hat R_{ab},d_{abc}{}^d,\Omega)$,
$\hat R_{ab}$ is the traceless part of the Ricci tensor, and
 $\Omega \, d_{abc}{}^d$ the Weyl tensor of $g_{ab}$.
Note that the equations are regular even for $\Omega=0$.

The 3+1 decomposition of the conformal geometry can be carried out as usual
in general relativity, e.g.
$$
g_{ab} = h_{ab} - n_{a} n_{b} = \Omega^2 (\tilde h_{ab} - \tilde n_{a}
                                                          \tilde n_{b}) \; ,
$$
where $h_{ab}$ and $\tilde h_{ab}$ are the Riemannian 3-metrics
induced by $g_{ab}$ respectively $\tilde g_{ab}$ on a spacelike hypersurface
with unit normals $n_a$, and equivalently $n_{a} = \Omega \,  \tilde n_{a}$
(our signature is $(-,+,+,+)$).
The relation of the extrinsic curvatures
($\tilde k_{ab}=\frac{1}{2}{\cal L}_{\tilde n}\tilde h_{ab}$,
$k_{ab}=\frac{1}{2} {\cal L}_{n} h_{ab}$) is then easily derived as
$k_{ab} = \Omega (\tilde k_{ab} + \Omega_0 \tilde h_{ab})$, where
$\Omega_0 = n^a \nabla_a \Omega$.
 
The additional variables $\hat R_{ab}$ and $d_{abc}^{d}$
can be decomposed into spatial objects by 
${}^{\scriptscriptstyle(0,1)\!}\hat R_a = n^{b}{h_a}^c \hat R_{bc}$,
${}^{\scriptscriptstyle(0,1)\!}\hat R_{ab} = {h_a}^c {h_b}^d \hat R_{bd}$,
$E_{ab} = d_{efcd}   h^e{}_a n^f h^c{}_b n^d$,
$B_{ab} = d^*_{efcd} h^e{}_a n^f h^c{}_b n^d$,
where $E_{ab}$ and $B_{ab}$ are called the electric and magnetic components
of the rescaled Weyl tensor ${d_{abc}}^{d}$.

Note that for regular components of
$h_{ab}$ and $k_{ab}$, the corresponding components of $\tilde h_{ab}$ and
$\tilde k_{ab}$
with respect to the same coordinate system will in general diverge 
due to the compactification effect. However for the coordinate independent
traces $k= h^{ab} k_{ab}$, $\tilde k = \tilde h^{ab} \tilde k_{ab}$
of the extrinsic curvatures we get
$$
\Omega k = (\tilde k + 3 \Omega_0) \; ,
$$
which can be assumed regular everywhere.
Note that at ${\mycal I}$, $\tilde k = - 3 \Omega_0$. Since ${\mycal I}^+$ is an ingoing
null surface (with $(\nabla_a \Omega) (\nabla^a \Omega) = 0$ but
$\nabla_a \Omega \neq 0$ ), we have that $\Omega_0 < 0$ at  ${\mycal I}^+$.
It follows that $\tilde k > 0$ at ${\mycal I}^+$.
We will thus call regular spacelike hypersurfaces in ${\cal M}$ hyperboloidal
hypersurfaces, since in $\tilde {\cal M}$ they are analogous to the standard
hyperboloids $t^2 - x^2 - y^2 - z^2 = {3\over\tilde k^2}$ in Minkowski space,
which provide the standard example.
Since such hypersurfaces cross ${\mycal I}$ but are
everywhere spacelike in ${\cal M}$, they allow to access ${\mycal I}$ and
radiation quantities defined there by solving a Cauchy problem (in contrast
to a characteristic initial value problem which utilizes a null surface
slicing). 
Note that in a globally hyperbolic physical spacetime,
hyperboloidal hypersurfaces will determine the future of the physical
spacetime, but not all of its past,
we therefore call our studies {\em semiglobal}.

The timelike vector $t^a = (\partial/\partial t)^a$ is decomposed
in the standard way
into a normal and a tangential component:
\begin{equation}\label{ta_split}
t^{a} = N n^{a} + N^{a},\quad N^{a}n_{a} = 0 \; .
\end{equation}
$N$ is called the lapse
function, because it determines how fast
the time evolution is pushed forward in the direction normal to $S$, and
thus determines ``how fast time elapses''.
The tangential component $N^{a}$, $N^{a}n_{a}=0$,
shifts spatial coordinate points with time evolution,
accordingly $N^{a}$ is called {\em shift} vector.
The lapse $N$ and shift $N^{a}$ are {\em not} dynamical quantities, they can
be specified freely and correspond to the arbitrary
choice of coordinates: the lapse determines the slicing of spacetime,
the choice of shift vector determines the spatial coordinates.

We will not discuss the full $3+1$ equations here for brevity, but rather refer
to \cite{PeterI}. Their most essential feature is that they
split into constraints plus symmetric hyperbolic evolution equations
\cite{PeterI}.
The evolution variables are
$h_{ab}$,
$k_{ab}$,
         the connection coefficients
$\gamma^a{}_{bc}$,
${}^{\scriptscriptstyle(0,1)\!}\hat R_a$,
${}^{\scriptscriptstyle(0,1)\!}\hat R_{ab}$,
$E_{ab}$,
$B_{ab}$, as well as
$\Omega$, $\Omega_0$, $\nabla_a \Omega$, $\nabla^a \nabla_a \Omega$
-- in total this makes $57$ quantities.
In addition the gauge source functions $q$, $R$ and $N^a$ have to be specified,
in order to guarantee symmetric hyperbolicity they are given as functions of
the coordinates.
Here $q$ determines the lapse 
as $N= e^q \sqrt {\det h}$ and $N^a$ is the shift vector. The Ricci scalar $R$
can be thought of as implicitly steering the conformal factor $\Omega$.

The constraints of the conformal field equations (see  (14) of 
\cite{PeterI}) are regular equations on the whole conformal spacetime
$({\cal M}, g_{ab})$, but they have not yet been cast into a standard type of
PDE system, such as a system of elliptic PDEs
(recently however, some progress in this direction has been achieved
by Butscher \cite{Adrian}). Therefore some remarks on how to proceed in this
situation are in order. A possible resolution is to
resort to a 3-step method \cite{AnCA92ot,PeterII,JoergIII}:
\begin{enumerate}
\item Obtain data for the Einstein equations: the first and
second fundamental forms $\tilde h_{ab}$ and ${\tilde k}_{ab}$ induced
on $\tilde\Sigma$ by $\tilde g_{ab}$, corresponding in
the compactified picture to $h_{ab}$, ${k}_{ab}$ and 
$\Omega$ and $\Omega_0$.
 This yields so-called ``minimal data''.
\item  Complete the minimal data on $\bar\Sigma$  to data for {\em all}
 variables using the conformal
constraints -- {\em in principle} this is mere algebra and
differentiation.
\item Extend the data from $\bar\Sigma$ to $\Sigma$ in some ad hoc but
sufficiently smooth and ``well-behaved'' way.
\end{enumerate}

In order to simplify the first step, numerical implementations \cite{PeterII,PeterIII,JoergIII} so far have been restricted to a subclass of
hyperboloidal slices where initially ${\tilde k}_{ab}$ is pure trace,
${\tilde k}_{ab} = \frac{1}{3} {\tilde h}_{ab} \tilde k$.
The momentum constraint
\begin{equation}\label{eq:MC}
  \tilde\nabla^b {\tilde k}_{ab} - \tilde\nabla_a \tilde k = 0  
\end{equation}
then implies $\tilde k = \mbox{const.} \ne 0$. We always set $\tilde k > 0$.
In order to reduce the Hamiltonian constraint
$$
\label{HamilConstr}
 {}^{\scriptscriptstyle (3)\!\!} \tilde R + {\tilde k}^2 
               = {\tilde k}_{ab}{\tilde k}^{ab}
$$
to {\em one} elliptic equation of second order,
we use a modified Lichnerowicz ansatz 
$$
  \label{tildeh}
  {\tilde h}_{ab} = \bar\Omega^{-2} \phi^4 h_{ab}
$$
with {\em two} conformal factors $\bar\Omega$ and $\phi$. The principal idea
is to choose  $h_{ab}$ and $\bar\Omega$, and solve for $\phi$, as we will
describe now.
First, the ``boundary defining'' function $\bar\Omega$ is chosen to
vanish on a 2-surface ${\cal S}$ -- the boundary of $\bar\Sigma$
and initial cut of ${\mycal I}$ -- with non-vanishing gradient on ${\cal S}$.
The topology of  ${\cal S}$ is chosen as spherical for asymptotically
Minkowski spacetimes. 
Then we choose $h_{ab}$ to be a Riemannian metric on $\Sigma$,
with the only restriction
that the  extrinsic 2-curvature induced by $h_{ab}$ on ${\cal S}$ is pure
trace, which is required as a smoothness condition \cite{AnCA92ot}. 
With this ansatz ${\tilde h}_{ab}$ is singular at ${\cal S}$,
indicating that ${\cal S}$ represents an infinity.
The Hamiltonian constraint then reduces to the Yamabe equation for the
conformal factor $\phi$:
$$
  4 \, \bar\Omega^2 \Delta\phi
  - 4 \, \bar\Omega (\!\nabla^a \bar\Omega)(\!\nabla_a \phi)
  - \left( \frac{1}{2} {}^{\scriptscriptstyle (3)\!\!}R \, \bar\Omega^2
  + 2 \bar\Omega \Delta\bar\Omega
           - 3 (\!\nabla^a \bar\Omega) (\!\nabla_a \bar\Omega) 
    \right) \phi
  = \frac{1}{3} {\tilde k}^2 \phi^5 \; .
$$
This is a semilinear elliptic equation -- except at ${\cal S}$,
where the principal part vanishes for a regular solution. This
however determines the boundary values as
\begin{equation}\label{boundaryvals}
\phi^4 
= \frac{9}{{\tilde k}^2} (\nabla^a \bar{\Omega}) (\nabla_a \bar{\Omega}) \; .
\end{equation}
Existence and uniqueness of a positive solution to the Yamabe 
equation and the corresponding existence and uniqueness of regular data for
the conformal field equations using the approach outlined above (assuming the
``pure trace smoothness condition) have been
proven by Andersson, Chru\'sciel and Friedrich~\cite{AnCA92ot}.

If the Yamabe equation is solved numerically, the boundary has to be chosen
at  ${\cal S}$, the initial cut of ${\mycal I}$,
with boundary values satisfying  (\ref{boundaryvals}).
If the equation would be solved on a larger grid
(conveniently chosen to be
Cartesian), boundary conditions would have to be invented, which
generically would cause the solution to lack sufficient
differentiability at ${\cal S}$,
see H\"ubner's discussion in \cite{PeterII}. This problem is due to
the degeneracy
of the Yamabe equation at $\cal S$. Unfortunately, this means
that we have to solve an elliptic problem with {\em spherical boundary}.

The constraints needed to complete minimal initial data to data for 
all evolution variables split into two groups: those that require divisions
by the conformal factor $\Omega$ to solve for the unknown variable,
and those which do not.
The latter do not cause any problems and can be solved without
taking special care at $\Omega = 0$.
The first group, needed to compute ${}^{\scriptscriptstyle(1,1)\!}\hat R$,
$E_{ab}$ and $B_{ab}$,
however does require special numerical
techniques to carry out the division,
and furthermore it is not known whether solving them on the whole
Cartesian time evolution grid
actually allows solutions which are sufficiently smooth
across ${\mycal I}$. Thus, at least for these we
have to find some ad-hoc extension. There are however also examples
of analytically known initial data, e.g. for the Minkowski and Kruskal
spacetimes, where all constraints are solved on the whole 
Cartesian time evolution grid.

\section{Examples: Different Ways to Compactify Minkowski Spacetime}
\label{sec:examples}
%%%%%%%%%%%%%%%%%%%%%%%%%%%%%%%%%%%%%%%%%%%%%%%%%%%%%%%%%%%%%%%%%%%%%%%%

The examples presented in this section
help to illustrate the compactification procedure -- in particular its
inherent gauge freedom, and they  yield interesting numerical tests, some
of which will be presented in Sec. \ref{sec:results}.

\subsection{Almost Static Compactification of Minkowski Spacetime}
\label{sec:MinkTextbook}
%%%%%%%%%%%%%%%%%%%%%%%%%%%%%%%%%%%%%%%%%%%%%%%%%%%%%%%%%%%%%%%%%%

From the perspective of hyperboloidal initial data, the simplest way to
compactify Minkowski spacetime is to
choose the initial conformal three-metric as the flat metric,
$h_{ab} = \delta_{ab}$, to set $k_{ab} = h_{ab}$, which solves
the momentum constraint  (\ref{eq:MC}), and to choose the conformal
curvature scalar $R_g$ \footnote{We change notation from
$R$ to  $R_g$ for this section to avoid confusion with a coordinate $R$ we
will introduce below.} as spherically symmetric, $R_g = R_g(x^2+y^2+z^2)$.
We know from \cite{AnCA92ot}
that a unique solution to the constraints exists, it is not hard to see that it
has to be spherically symmetric. Furthermore, it is topologically trivial.
From Birkhoff's theorem we can thus conclude that we are dealing with
Minkowski spacetime.
Choosing the simplest gauge $q=0$, $N^a = 0$, $R_g=0$, the resulting unphysical
spacetime is actually Minkowski spacetime in standard coordinates:
$$
ds^2 = -\D t + \D \Sigma^2  = \Omega^2(-\D T^2 + \D R^2 +
R^2 \left(\D \theta^2+\sin^2\theta \D \phi^2\right) \; ,
$$
where $d\Sigma^2$ is the standard metric on $R^3$,
$d\Sigma^2=dr^2 + r^2 \left(d\theta^2+\sin^2\theta d\phi^2\right)$,
and the conformal factor is
\begin{equation}\label{eq:OmegaMinkMink}
 \Omega = \left( R^2 - T^2 \right)^{-1} = \left( r^2 - t^2 \right) \; ,
\end{equation}
where
$$
r  =  \frac{R}{R^2 - T^2}\; , \qquad
t  =  \frac{T}{R^2 - T^2}\; .
$$
This setup has been chosen as the basis of H\"ubner's numerical study of weak
data evolutions \cite{PeterIV}. With the initial cut of ${\mycal I}^+$ chosen
at $x^2 + y^2 + z^2 = 1$, $i^+$ is located at coordinate time $t=1$,
the generators of ${\mycal I}^+$ being straight lines at an angle of $45^\circ$.

This conformal representation of Minkowski spacetime constitutes an
``almost static'' gauge -- since the spatial geometry is time-independent,
so are all evolution variables {\em except}
for the conformal factor $\Omega$. The physical region inside of ${\mycal I}^+$
contracts to the regular point $i^+$ within finite time.
This feature is shared with the standard ``textbook'' example of conformally
compactifying Minkowski spacetime, which takes the form of a map into part of
the Einstein static universe with $R_g = 6$,
\begin{equation}\label{eq:StandardGauge}
\D s^2 = -\D t^2 + \D \Sigma^2  = \Omega^2 \, \left(-\D T^2 + \D R^2 +
R^2 \left(\D \theta^2 + \sin^2\theta \D \phi^2\right)\right) \; ,
\end{equation}
where $d\Sigma^2$ is the standard metric on $S^3$,
$d\Sigma^2=d\varrho^2 + \sin^2 \varrho \left(d\theta^2+\sin^2\theta d\phi^2\right)$,
and the conformal factor is
$$
 \Omega^2 =4 \, (1 + (T-R)^2)^{-1} \, (1 + (T+R)^2)^{-1} =
  4 \cos^2 \frac{t-\varrho}{2} \cos^2 \frac{t+\varrho}{2} \; .
$$
Here the coordinate transformations are
\begin{eqnarray}
\varrho &=& \arctan (T+R) - \arctan (T-R)\; , \\
t       &=& \arctan (T+R) + \arctan (T-R)\; .
\end{eqnarray}
In these coordinates Minkowski spacetime corresponds to the coordinate
ranges
\begin{eqnarray}
  - \pi &<&    t + \varrho < \pi \; , \\
  - \pi &<&    t - \varrho < \pi \; , \\
\varrho &\geq& 0 \; .
\end{eqnarray}
For details and pictures of this mapping see the discussions by
\cite{Pe63ap}, \cite{Wald} or \cite{Stewart}.

Alternatively, we can choose stereographic spatial coordinates such that
$$
\D \Sigma^2 =
 \omega^2\left(\D r^2 + r^2\left(\D \theta^2
 +\sin^2\theta \D \phi^2\right)\right) \; ,
\qquad \omega = 2/(1 + r^2) \; .
$$
or we may absorb the spatial conformal factor into the spacetime conformal
factor by rescaling to
\begin{equation}\label{eq:pseudostatic_spatiallyflat}
\D s'^2 =  -\omega^{-2} \D t^2 + \D r^2 + r^2
          \left(\D \theta^2+\sin^2\theta \D \phi^2\right) \; ,
\end{equation}
which yields the lapse to be $N=1$ ($q = -3 \log{\omega}$),
respectively $N=\omega^{-2}$ ($q = -2 \log{\omega}$).
Note that in the numerical code we use
Cartesian coordinates  $x = r \sin \theta \cos \phi$,
$y = r \sin \theta \sin \phi$, $z = r \cos \theta$.

The conformal transformation leading to  
(\ref{eq:pseudostatic_spatiallyflat}) changes the scalar curvature
from $R_g = 6$ to $R_g = -12 \, (1+r^2)^{-2}$.
We will see below in Sec. \ref{sec:results} that these simple variations in
gauge source functions and conformal rescaling lead to numerical
representations which are quite different, e.g. with regard to accuracy and
robustness.

\subsection{A Static Hyperboloidal Gauge for Minkowski Spacetime}
\label{sec:MinkStat}
%%%%%%%%%%%%%%%%%%%%%%%%%%%%%%%%%%%%%%%%%%%%%%%%%%%%%%%%%%%%%%%%%%%%

By translating a standard hyperboloid in Minkowski
spacetime along the trajectories of the $\partial/\partial t$ Killing vector,
one can obtain a gauge where not only the conformal spacetime is static, but
also the conformal factor is time-independent -- thus also the physical
geometry and all evolution variables of the conformal field equations can be
made time independent
(this has been pointed out me by M. Weaver, and I essentially
follow her notes below). See also a talk given by V. Moncrief
\cite{talk_Moncrief},
that we have become aware of after starting to work with this gauge.

In this gauge the point $i^+$ is not brought into a finite distance and
remains in the infinite future.
This conformal gauge is particularly useful for stability tests.

To derive this static metric we start
with spherical coordinates $(T,R,\theta,\phi)$ on Minkowski space,
where the metric is
\begin{equation}
\D \tilde s^2 = -\D T^2
+ \D R^2 + R^2 \left( \D \theta^2 + \sin^2\theta \D \phi^2\right) \; .
\end{equation}
A family of standard hyperboloids with time translation parameter $t$
is given by
$$
(T-t)^2 - R^2 = 1 \; . 
$$
We transform now to new coordinates $(t,\varrho,\theta,\phi)$,
where the level surfaces of $t$ are the standard hyperboloids, and
$\varrho(R)$ is chosen as a new radial parameter on the  hyperboloids. Setting 
$T = t +  \cosh \varrho$ and $R = \sinh \varrho$,
the physical metric becomes
\begin{equation}
\D \tilde s^2 = -\D t^2 -2 \sinh \varrho \, \D \varrho \, \D t + \D \varrho^2
+ \sinh^2 \varrho \left(\D \theta^2 + \sin^2\theta \D \phi^2\right) \; .
\end{equation}
For simplicity we choose the conformal three-metric to be flat and
introduce new spherical coordinates $(r, \theta, \phi)$, such that
\begin{equation}
\D s^2\vert_{t=const.}
    = \D r^2 + r^2 \left(\D \theta^2 + \sin^2\theta \D \phi^2\right).
\end{equation}
Since $h_{ab} = \Omega^2 \tilde h_{ab}$ we get $\Omega = dr/d\varrho$ and
\begin{equation}
\int_\varrho^\infty {\D \varrho' \over \sinh \varrho} 
   = \int_r^1 {\D r' \over r'} \; .
\end{equation}
The limits of integration are given by the fact that
$\lim_{\varrho \to \infty} r = 1$.
Performing the integrals one finds that
\begin{equation}
r = {e^\varrho - 1 \over e^\varrho + 1} = {1 \over R} (\sqrt{1 + R^2} -1)
\end{equation}
and
\begin{equation}
\Omega = \frac{1 - r^2}{2} \; ,
\end{equation}
our choice thus maps ${\mycal I}^+$ to the timelike cylinder $r=1$.

The computer time coordinate $t$ is a Bondi time coordinate
on ${\mycal I}$. 
In coordinates $(t,r,\theta,\phi)$ the conformal metric reads
\begin{equation}
\D s^2 = -\Omega^2 \D t^2 - 2 \, r \, \D r \, \D t  + \D r^2 
         + r^2 \left(\D \theta^2 + \sin^2\theta \D \phi^2\right) \; ,
\end{equation}
or
\begin{equation}\label{eq:static_gauge}
\D s^2 =
-\Omega^2 \D t^2 - 2 \, \D t \, \left(x \, \D x + y \, \D y + z \, \D z\right) 
 + \D x^2 + \D y^2 + \D z^2 
\end{equation}
in Cartesian coordinates  $x = r \sin \theta \cos \phi$,
$y = r \sin \theta \sin \phi$, $z = r \cos \theta$,
which are used in the numerical code.
The shift vector is thus given by $N^i = -x^i$, and the lapse can be computed
from $-N^2 + h_{a b} N^a N^b = g_{t t}$ (as implied by  (\ref{ta_split})) as
$N = (1 + r^2)/2$. The three metric has unit determinant,
so $q = \ln N$. Note that the shift vector does not become ``superluminal''
beyond ${\mycal I}^+$, because the lapse is growing faster than the shift --
$g_{tt}$ is nonnegative everywhere, and zero only at  ${\mycal I}^+$.
The conformal Ricci scalar is
\begin{equation}
R_g = 12 { (1 - r^2) (3 + r^2) \over (1 + r^2)^3} \; ,
\end{equation}
which vanishes at ${\mycal I}^+$.

For a numerical calculation one needs the minimal initial data set,
\begin{equation}
(h_{ab}, \Omega, k_{ab}, \Omega_0)
\end{equation}
and the gauge source functions, $(R, N, N^a)$.  In a numerical
calculation in which the Yamabe equation is solved to find
$\Omega$, one gives $(h_{ab}, \bar \Omega, \mbox{tr}\,k)$.
In a test case such as this which is an explicitly known solution,
one can just take $\bar \Omega = \Omega$.  It remains therefore
to  calculate $k_{ab}$ and $\Omega_0$.  From
\begin{equation}
k_{ab} = {1 \over 2 N} (\partial_t h_{ab} - {\cal L}_{N}h_{ab})
\end{equation}
we find that the components of the extrinsic curvature are
$k_{ij} = {1 \over N} \delta_{ij}$
and $k = {3 \over N}$.  From the identity 
$\tilde k = \Omega \, k - 3 \, \Omega_0$ we find that
\begin{equation}
\Omega_0 = -\frac{2 r^2 }{1 + r^2} \; .
\end{equation}

\section{History}\label{sec:history}
%%%%%%%%%%%%%%%%%

This section tries to give a broad overview of what has been achieved
so far in the field of numerical treatment of the conformal field equations.
Historically, this field was started by Peter H\"ubner by studying
a scalar field coupled to gravity in spherical symmetry in his PhD thesis
\cite{Hu93nu} finished in 1993. His subsequent work has lead to the development
of both a 2D and a 3D evolution code, formulated in ``metric'' variables.
J\"org Frauendiener has also developed an independent 2D  code, formulated
in frame variables. 

\subsection{Early work in spherical symmetry}\label{sec:1Dcodes}
%%%%%%%%%%%%%%%%%%%%%%%%%%%%%%%%%%%%%%%%%%%%%%%%%%%%%%%%%%%%%%%%%%%%

The first numerical implementation of the conformal field equations
is due to Peter  H\"ubner, who has studied the spherically symmetric
collapse of scalar fields in his PhD thesis \cite{Hu93nu}, and
subsequently in \cite{Hu96mf}.
In his gauge both future null infinity (${\mycal I}{}^+$) and future timelike
infinity ($i^+$) are compactified, and the whole spacetime is covered in finite
coordinate time.
H\"ubner studies the global structure
of the spacetime, including the appearance of singularities and the
localization of the event horizon. To handle the
latter, floating point exceptions are caught and
grid points are flagged as ``singular'', grid points whose values depend
on information from singular grid points are correspondingly flagged as
singular as well. Even though this method does not allow to
actually trace the singularity in a strict sense
(computers can not actually deal with infinite values), 
the method traces the singularity as tightly as possible.
In contrast to typical black hole excision schemes, which
are based on locating the apparent horizon, this scheme could thus be
termed ``tight excision''.  The method has not yet
been implemented in higher dimensions, where one has to face more intricate
technical problems, and where the structure of the singularity is likely to
be much more complex as well.
The paper also studies critical collapse. H\"ubner's results are
consistent with the black hole mass power-law scaling with the
correct exponent, however no echoing related to discrete self-similarity
has been seen in his results. This has created some discussion, whether
the results of other authors are numerical artifacts, or artefact's
of boundary conditions at finite distance, however numerical
critical collapse simulations in a compactified characteristic framework
have recently
shown both the correct power-law scaling and discretely
self-similar echoing \cite{michael_puerrer}.

The coordinates in this approach are based on the geometric
structure of double null-coordinates available in spherical symmetry.
Unfortunately this choice does not generalize in the absence of
spherical symmetry. Finding a gauge that would allow to run, say,
the Kruskal spacetime in a 3D code for ``arbitrarily long'' Bondi times,
is an open problem, where significant insight could be gained
from studying more general gauges in a manifestly spherically symmetric
code.

\subsection{Axially symmetric spacetimes with toroidal Scri in the frame formulation}\label{sec:2Dcodes}
%%%%%%%%%%%%%%%%%%%%%%%%%%%%%%%%%%%%%%%%%%%%%%%%%%%%%%%%%%%%%%%%%%%%%%%%%%

Following H\"ubner's encouraging results for spherically symmetric simulations
\cite{Hu93nu,Hu96mf}, numerical
codes have been developed by Frauendiener and H\"ubner to
study axially symmetric spacetimes. For simplicity, e.g. to avoid
numerical stability problems at the axis of symmetry, and to avoid
problems associated with a ${\mycal I}$
of spherical topology -- which does not align with Cartesian coordinates --
both  H\"ubner and Frauendiener considered the asymptotically A3-spacetimes
\cite{schmidt-a3,foertsch}, which do not possess an axis of symmetry and
where ${\mycal I}$ has toroidal topology. These spacetimes are modeled after the 
A3-metric in the Ehlers-Kundt classification \cite{Ehlers_Kundt},
which provides
an analogue of the Schwarzschild metric in plane symmetry. 
These spacetimes are not physical,
but they contain a large class of nontrivial radiative vacuum
spacetimes, which make them an interesting toy model to study numerical
techniques, gauges, and the extraction of radiation \footnote{One of
the notable differences to the Minkowski case is that one can only define a
Bondi-mass but no Bondi four-momentum.}
These axisymmetric codes thus have been designed to treat the vacuum case,
and matter couplings have not yet been implemented.
An advantage for code-testing is that exact solutions are known
\cite{schmidt-a3,PeterA3}.

In the first \cite{JoergI} of a series of papers \cite{JoergI,JoergII,JoergIII,JoergIV} on his axisymmetric
code, Frauendiener gives a nice overview of the motivation for using the
conformal field equations of numerical simulations of isolated systems.
He discusses the conformal field equations in the space spinor formalism
\cite{space-spinors}, which is chosen because of compactness of notation,
and because it allows a very straightforward 3+1 decomposition of the
equations, rendering the equations in symmetric hyperbolic form.
His formalism contains 8 free functions
which determine the gauge: the harmonicity $F := \nabla_c \nabla^c t$
determines the choice of time coordinate $t$, the shift is given in terms
of frame coefficients, the scalar curvature $R$ ($\Lambda$ in his notation)
of the compactified spacetime, and an imaginary and symmetric space 
spinor field $F_{AB}$ (i.e. three numbers) which determines rotations of
the spatial frame (for $F_{AB}=0$ the frame is transported via Fermi-Walker
transport).
He also discusses the implications of the assumptions of the toroidal
symmetry, in particular for the choice of gauge -- e.g. the adaption
of the frame.

In the second paper \cite{JoergII} of the series
Frauendiener discusses his numerical methods and gauge choices, and presents
results
from evolutions of initial data corresponding to the exact solution
presented in \cite{schmidt-a3}. Here one of the two Killing vectors is
disguised by a coordinate transformation.
The numerical evolution proceeds via a generalization of the Lax-Wendroff
scheme to 2D, which Frauendiener proves to be stable and
second order accurate. The time step is such that the numerical
domain of dependence is contained in the domain of dependence as defined
by the equations.
An essential difficulty -- as usual -- is posed by the treatment of the 
boundary.
Well-posedness of the associated initial-boundary-value
problem has not yet been proven, and numerical analysis can only provide
rough guidelines to work out stable algorithms \cite{Trefethen}.
Frauendiener's boundary treatment is based on the identification of ingoing
and outgoing modes at the boundary, as determined from the symmetric hyperbolic
character of the equations.
He sets boundary values for inward-propagating quantities (e.g. motivated
by the exact solution), and sets values for the outward propagating quantities
by extrapolation from the interior.
This method can be applied just a few grid-points outside of ${\mycal I}$ and is
found to be stable as long as the gauge source functions dot not depend
on the evolution variables -- which would change the characteristics.
Note that the constraints will in general not be satisfied on the boundary,
which may trigger constraint-violating modes of the equations.

Frauendiener gives a detailed discussion of the problems associated with the
choice of the gauge, and performs a number of numerical
experiments in this respect, evolving data corresponding to exact solutions
\cite{schmidt-a3,PeterA3} with singular $i^+$.
One of the problems is, that if the gauge source functions are allowed
to depend on the evolution variables, this will change the characteristics
of the system and will in general spoil the symmetric hyperbolic
character of the system. Experiments in this direction, where
$F=F(N,K)$ indeed exhibited a boundary instability.
Regarding the choice of time coordinate, that is the harmonicity
function $F$, several choices are tested: a ``natural'' gauge, which is taken from the exact solution, the Gauss gauge (where the lapse $N$ is spatially
constant), the harmonic gauge, $F=0$, and a family of gauges that
interpolates between the ``natural'' and harmonic gauge.

The ``natural'' gauge is found to provide the best performance and the
approach to the singularity is found to be 
essentially limited by machine precision.
The harmonic gauge leads to a coordinate singularity before reaching
the singularity, this feature is shared by most of the
gauges that interpolate between natural and harmonic gauge.
For the ``Gauss'' gauge with $N=const.$  caustics (coordinate shocks)
develop quickly and crash the simulations.

Regarding the choice of shift vector, a prescription for ${\mycal I}$ fixing -- that
is steering the evolution of the surface $\Omega=0$ is discussed, which
can be easily implemented in Frauendiener's formulation. This however relies on the specific form of the frame equations, and does not carry over to
to equations as used in H\"ubner's codes \cite{PeterI}.
In particular he studies the case of
``${\mycal I}$ freezing'' -- holding the coordinate position of ${\mycal I}$ in
place such that no loss of resolution occurs in the physical domain.

Finally, he discusses the extraction of gravitational radiation, e.g. by
computing the Bondi mass, and shows some results.

In order to study more general spacetimes, Frauendiener has implemented
a numerical scheme for determining hyperboloidal initial data
sets for the conformal field equations
by using pseudo-spectral methods as described in \cite{JoergIII}.
He uses the implicit approach of first solving the Yamabe equation,
and then carrying out the division by the conformal factor for
certain fields which vanish on ${\mycal I}$. The challenge there is to
numerically obtain a smooth quotient. The division problem is treated by
a transformation to the coefficient space, where a
QR-factorization of a suitable matrix is used, and then transforming back. 

In \cite{JoergIV} Frauendiener gives a pedagogical discussion
of the issue of radiation extraction in
asymptotically flat space-times within the framework of conformal methods for
numerical relativity. The aim is to show that there exists a well defined and
accurate extraction procedure which mimics the physical
measurement process and operates entirely intrinsically within ${\mycal I}^+$.
The notion of a detector at infinity is defined by idealizing local
observers in Minkowski space. A detailed discussion is presented for Maxwell
fields and the generalization to linearized and full gravity
is performed by way of the similar structure of the asymptotic fields.

Recently, Hein has written a 2D axisymmetric code that allows for an axis
\cite{Hein-diploma}, i.e. can treat the physical situation with a ${\mycal I}^+$ of
spherical topology. The usual problem of the
coordinate singularity at the axis in adapted coordinates was
solved by using Cartesian coordinates, following a method developed by
Alcubierre et al. \cite{cartoon}.
The code has so far been tested by evolving Minkowski
spacetime in various gauges, further tests with nontrivial
spacetimes are currently underway.

\subsection{Metric-based 2D and 3D codes}\label{sec:3Dcode}
%%%%%%%%%%%%%%%%%%%%%%%%%%%%%%%%%%%%%%%%%%%%%%%%%%%%%%%%%%%%%%%%%%%%%%%%

The basic design of H\"ubner's approach is outlined in \cite{PeterI},
where he presents the first order time evolution equations as obtained from
a 3+1 split of the conformal field equations.
The evolution equations can be brought into symmetric hyperbolic form
by a change of variables. 
He discusses his motivation of avoiding artificial boundaries and how
the conformal field  equations formally allow placement of the grid boundaries
outside the physical spacetime.

A particularly subtle part of the evolution usually is the boundary treatment. 
In the conformal approach we are in the situation that the boundary can
actually be placed outside of the physical region of the grid -- this is one of
its essential advantages! In typical explicit time evolution algorithms,
such as our Runge-Kutta method of lines, the numerical propagation speed
is larger than the speed of all the characteristics (in our case
the speed of light). Thus ${\mycal I}$ does {\em not} shield the physical region
from the influence of the boundary  -- but this influence has to converge
to zero with the convergence order of the algorithm --
fourth order in our case. In principle one therefore does not have to
choose a ``physical''
boundary condition, the only requirements are stability and ``practicality'' --
e.g. the boundary condition should avoid, if possible, the development of
large gradients in the unphysical region to reduce the numerical 
``spill over'' into the physical region, or even code crashes.
It seems likely however, that this practicality requirement will
eventually lead to a treatment of the boundary which satisfies the
constraints at the boundary.

H\"ubner develops the idea of modifying the
equations near the grid boundaries to obtain
a consistent and stable discretization.
The current implementation of the boundary treatment
relies on this introduction of a ``transition layer''
in the unphysical region, which is used to transform  the rescaled
 Einstein equations to trivial evolution equations, which are stable with
a trivial copy operation at outermost gridpoint as a boundary condition
(see  \cite{PeterI} for details and references).
He thus replaces
$$
  {  \partial_t {f} 
  + {{A}}^i \partial_i {f} - 
  {b} = 0}
$$
by
$$
  { \partial_t {f} + \alpha(\Omega) \left( {{A}}^i \partial_i {f} -
  {b} \right) = 0,}
$$
where $\alpha$ is chosen as $\alpha(\Omega) = 0$ for
$\Omega\leq\Omega_0<\Omega_1<0$ and $1$ for
$\Omega\geq\Omega_1$.
One potential problem is that the region of large constraint violations outside
of ${\mycal I}$ may trigger constraint violating modes of the equations that can
grow exponentially. Another problem is that a ``thin'' transition zone causes
large gradients in the coefficients of the equations -- thus eventually leading
to large gradients in the solution, while a ``thick'' transition zone
means to loose many gridpoints. If no transition zone is used at all, and the
Cartesian grid boundary touches ${\mycal I}$, the ratio of the number of 
grid points in the physical region versus the number of 
grid points in the physical region is already $\pi/6 \approx 0.52$.

Furthermore he discusses his point of view
concerning possible advantages of the conformal approach
and discusses potential problems of the Cauchy and Cauchy-Characteristic
matching approaches to numerical relativity. He outlines the geometric
scenario of his approach and stresses that these techniques allow,
in principle, to calculate the complete future of scenarios such as initial
data for $N$ black holes.

The second paper \cite{PeterII} of the series deals with the technical
details of construction of initial data and of the time-evolution
of such data.
The second and fourth order discretizations, which are used for the
construction of the complete data set and for the numerical integration of
the time evolution equations, are described and their efficiencies compared.
Results from tests for $A3$ and disguised Minkowski
spacetimes confirm convergence for the 2D and 3D codes.

The simplest approach to the division by $\Omega$ would be an implementation of
l'Hospital's rule, however this leads to nonsmooth errors and consequently
to a loss of convergence \cite{PeterII}.
Instead H\"ubner \cite{PeterII} has developed a technique to replace
a division
$g = f/\Omega$ by solving an elliptic equation of the type
$$
   \nabla^a \nabla_a ( \Omega^2 g - \Omega f ) = 0 
$$
for $g$ (actually
some additional terms added for technical reasons are omitted here for
simplicity).
For the boundary values $\Omega^2 g - \Omega f = 0$, the unique solution is
$g=f/\Omega$.
The resulting linear elliptic equations for $g$ are solved by the same 
numerical techniques as  the Yamabe equation.
For technical details see H\"ubner \cite{PeterIII}.

Finally, we have to extend the initial data to the full Cartesian
spatial grid in some way.
Since solving all constraints also outside of ${\mycal I}$ will in general not be
possible in a sufficiently smooth way \cite{PeterII},
we have to find an ad hoc extension,
which violates the constraints outside of ${\mycal I}$ but is sufficiently well
behaved to serve as initial data. The resulting constraint violation is not
necessarily harmful for the evolution, since ${\mycal I}$ causally disconnects
the physical region from the region of constraint violation. On the numerical
level, errors from the constraint violating region {\em will} in general
propagate into the physical region, but if our scheme is
consistent, these errors have to converge
to zero with the convergence order of the numerical scheme
(fourth order in our case). There may of course still be practical problems
that prevent us from reaching this aim: making the ad-hoc extension
well behaved is actually quite difficult, the initial constraint violation may
trigger constraint violating modes in the equations, which take us away from 
the true solution, singularities may form in the unphysical region, etc.

Since the time evolution grid is Cartesian, its grid points will in general not
coincide with the collocation points of the pseudo-spectral grid. Thus
fast Fourier transformations can not be used for transformation to
the time evolution grid. The current implementation instead uses
standard discrete (``slow'') Fourier transformations, which typically take up
the major part of the
computational effort of producing initial data. 

It turns out, that the combined procedure works reasonably well for 
certain data sets. For other data sets the division by $\Omega$
is not yet solved in a satisfactory way, and constraint violations
are of order unity for the highest available resolutions.
In particular this concerns the constraint
$\nabla_b E_a{}^b = - {}^{\scriptscriptstyle(3)\!\!}
                       \varepsilon_{abc} k^{bd} B_d{}^c$ ((14d) in
 \cite{PeterI}),
since $E_{ab}$ is computed last in the hierarchy of variables and
requires two divisions by $\Omega$. Further research
is required to analyze the problems and either improve the current 
implementation or apply alternative algorithms.
Ultimately, it seems desirable to change the algorithm of obtaining
initial data to a method that solves the conformal constraints directly and
therefore does not suffer from the current problems.
This approach may of course introduce new problems like an elliptic system
too large to be handled in practice.

The time evolution algorithm is an implementation of a standard
fourth order method of lines (see e.g. \cite{GuKA95TD}).
In the method of lines we formally write
\begin{equation}
  \partial_t f = B(f,\partial_i f) \; ,
\end{equation}
where $B(f,\partial_i{f}) = - A^i(f) \partial_i f + b(f)$.
Discretizing the spatial derivatives parameterizes the
ordinary differential equations by grid point index.
For the present code, fourth order accurate centered spatial differences 
have been implemented, e.g. for the $x$-derivative as:
$$
\partial_x f \rightarrow \frac{1}{12 \Delta x}\left(
-f_{i+2,j,k} + 8 f_{i+1,j,k} - 8 f_{i-1,j,k} + f_{i-1,j,k} \right) \; .
$$

The numerical integration of the ordinary differential equations
proceeds via the standard fourth order Runge-Kutta scheme:
\begin{eqnarray}
\label{RK4}
 f^{l+1}_{i,j,k} & = &
 f^{l}_{i,j,k}
  + \frac{1}{6}
    \left( k^l_{i,j,k}
           + 2 k^{l+1/4}_{i,j,k}
           + 2 k^{l+1/2}_{i,j,k}
           + k^{l+3/4}_{i,j,k}   \right) \; ,
\end{eqnarray}
where
\begin{eqnarray*}
  k^l_{i,j,k} & = &
  \Delta t \>
  {B}({f}^{l}_{i,j,k},\partial_i{f}^{l}_{i,j,k})\; ,
\\
  k^{l+1/4}_{i,j,k} & = &
  \Delta t \>
  {B}({f}^{l+1/4}_{i,j,k},
                \partial_i{f}^{l+1/4}_{i,j,k})\; , \quad
  {f}^{l+1/4}_{i,j,k} =
    {f^{l}_{i,j,k}} + \frac{1}{2} \, k^l_{i,j,k}\; ,
\\
  k^{l+1/2}_{i,j,k} & = &
  \Delta t \>
  {B}({f}^{l+1/2}_{i,j,k},
                \partial_i{f}^{l+1/2}_{i,j,k})\; , \quad
  {f}^{l+1/2}_{i,j,k} =
    {f^{l}_{i,j,k}}
    + \frac{1}{2} \, k^{l+1/4}_{i,j,k}\; ,
\\
  k^{l+3/4}_{i,j,k} & = &
  \Delta t \>
  {B}({f}^{l+3/4}_{i,j,k},
                \partial_i{f}^{l+3/4}_{i,j,k})\; , \quad
  {f}^{l+3/4}_{i,j,k} =
    {f^{l}_{i,j,k}} + k^{l+1/2}_{i,j,k}\; .
\end{eqnarray*}
Additionally, a dissipation term of the
type discussed in theorems ~6.7.1 and ~6.7.2 of  Gustafsson, Kreiss and
Oliger \cite{GuKA95TD} is added to the right-hand-sides to damp out high
frequency oscillations and keep the code numerically stable.  
The dissipation term used is
$\sigma Q_2 := \frac{\sigma }{64 \, N} (\Delta x)^5 \sum_{i=1}^{N}\partial_i{}^6 f$,
where the spatial derivatives are discretized as
\begin{eqnarray}
  \lefteqn{
    \partial_x{}^6 {f}^l_{i,j,k} \rightarrow
      \frac{1}{(\Delta x)^6}
      \left( {f}^l_{i-3,j,k} - 6 {f}^l_{i-2,j,k}
             + 15 {f}^l_{i-1,j,k} \right. }
 \nonumber\\
  & & \left. {}
             - 20 {f}^l_{i,j,k}
             + 15 {f}^l_{i+1,j,k} 
             - 6 {f}^l_{i+2,j,k} + {f}^l_{i+3,j,k}
    \right)\; . \nonumber        
\end{eqnarray}
Numerical experiments show that usually small amounts of dissipation 
($\sigma$ of order unity or smaller) are
sufficient, and do not change the results in any significant manner.
Numerical tests for Minkowski spacetime with disguised symmetries and
an explicitly known A3-like solution with radiation \cite{PeterA3}
are described in \cite{PeterII}.

Further extensive tests of the 2D code have been performed by Weaver
\cite{weaver-private_comm}.
She studied the choice of gauge source functions for an
A3-like solution, solving the Yamabe equation for the conformal factor.
She found that for this solution it is quite simple to prescribe
a shift so that ${\mycal I}$ is fixed to a very good approximation.
She also studied use of the gauge source function $q$ to prolong
the numerical simulation inside physical spacetime.  In cases where $q=0$
results in a ''singularity`` developing outside physical spacetime (which
causes the code to crash), prescription of $q$ so that the evolution
inside physical spacetime is prolonged
compared to outside allows the simulation to essentially
cover the physical spacetime to the future of the initial data surface.
She thus found that in this context the ad hoc prescription
of gauge source functions was sufficient to achieve desired effects,
and caused no instabilities.
Also she explored the effect of turning off the
transition zone, while still simply copying data at the outer
grid boundary into the ghost zone, along with prescription of
$q$ so that the evolution is slowed down at the outer boundary.
In the A3-like 2D runs this alternative boundary treatment was
successful, and avoided problems created by the transition zone.

In the third part of the series \cite{PeterIII}, a pseudospectral
solver for the constraints is described.
Since the implementation depends on the topology,
it discusses both the asymptotically A3 and asymptotically Minkowski cases.
At the end also some remarks are made about a possible extension to
the multi-black-hole case, using a multi-patch scheme (the Schwarz alternating
procedure).

In the fourth part of the series \cite{PeterIV} H\"ubner presents results of
3D calculations for initial data 
which evolve into a regular point $i^+$, and which thus could be called 
``weak data''.
The initial conformal metric is chosen in Cartesian coordinates as
\begin{equation}\label{eq:standard-h}
\D s^2 =  \left(1 + \frac{A}{3}
   \bar\Omega^2 \left( x^2 + 2 y^2 \right)\right) \D x^2 + \D y^2 + \D z^2 \; .
\end{equation}
The boundary defining function $\bar\Omega$
appearing in this ansatz is chosen as
$\bar\Omega = \left( 1 - \left(x^2+y^2+z^2\right) \right)/2$,
it is used to satisfy the smoothness
condition for the conformal metric at ${\mycal I}$.
For the gauge source functions H\"ubner has made the ``trivial'' choice:
$R=0$, $N^a = 0$, $q=0$, i.e. the conformal spacetime has vanishing scalar
curvature, the shift vanishes and the lapse is given by
$ N = e^{q} \sqrt{\det h} =\sqrt{\det h}$.
This simplest choice of gauge is completely sufficient for $A=1$ data, and has
lead to a milestone result of the conformal approach --
the evolution of weak data which evolve into a  regular point $i^+$
of $\cal M$, which is resolved as a single grid cell.
With this result H\"ubner has illustrated a theorem by Friedrich,
who has shown that for sufficiently weak initial data there exists a 
regular point $i^+$ of $\cal M$ \cite{Fr86ot}.
The complete future of (the physical part of) the initial slice
can thus be reconstructed in a finite number of computational time steps.
This calculation is an example of a situation for
which the usage of the conformal field equations is ideally suited:
main difficulties of the problem are directly addressed and solved
by using the conformal field equations.

A natural next question to ask is: what happens if one increases the
amplitude $A$? To answer this question, I have performed and analyzed
runs for integer values of $A$ up to $A=20$, preliminary results
have been presented in \cite{my_tuebingen_LNP}.
While for $A=1, 2$ the code was found to be able to continue
beyond $i^+$ without problems, for all higher
amplitudes the ``trivial'' gauge leads to code crashes before reaching $i^+$.
While the physical data still decay quickly in time, a sharp peak of the lapse
develops outside of ${\mycal I}$ and crashes the code after
Bondi time $\sim 8 (320 M)$ for $A=3$ and $\sim 1.5 (3 M)$ for $A=20$
(here $M$ is the initial Bondi mass).
A partial cure of the problem was obtained using a modified gauge source
function $q = -r^2/a$ ($N=e^{-r^2/a}\sqrt{\det h}$), where $a$ is tuned
such that one gets a smooth lapse and smooth metric components.
For, $A=5$, for example, a value of $a=1$ was found by moderate tuning of $a$ 
(significantly decreasing or increasing $a$ crashes the code before the
regular $i^+$ is reached).
Unfortunately, this modification of the lapse is not sufficient to achieve
much higher amplitudes. As $A$ is increased, the parameter $a$ requires
more fine tuning, which was only achieved for $A \leq 8$. For
higher amplitudes the code crashes with significant
differences in the maximal and minimal Bondi time achieved, while
the radiation still decays very rapidly. Furthermore, the curvature
quantities do not show excessive growth -- it is thus natural to assume that
we are still in the weak-field regime, and the crash is not connected to
the formation of an apparent horizon or singularity.
While some
improvement is obviously possible through simple non-trivial models for
the lapse (or other gauge source functions), this approach seems quite limited
and more understanding will be necessary to find practicable gauges.
An interesting line of research would be to follow the lines
of \cite{Miguel} in order to find evolution equations for the gauge source
functions which avoid the development of pathologies.

Schmidt has presented hyperboloidal initial data for the
Kruskal spacetime, a hyperboloidal foliation
for the future of these hyperboloidal initial data 
\cite{bernd_kruskal} and results from numerical simulations evolving
these initial data with different gauges, which have been performed by Weaver
with H\"ubner's 3D code. The explicit hyperboloidal version of the Kruskal
spacetime is very useful for numerically testing the conformal approach
in the treatment of black hole spacetimes. 
These runs have been performed in octant mode. The runs typically
proceed until the determinant of the three metric becomes negative
\cite{weaver-private_comm}, caused by some
feature in the exact solution which is no longer adequately resolved and
which is growing, leading to large narrow spikes
in the numerical data. Future work will have to be directed toward improving
the choice of gauge source functions such that rapidly growing sharp
features are avoided.

In the next section, I will present new results
obtained with the 3D code for asymptotically Minkowski spacetimes, which will
illustrate some of the current problems. One of these is the presence of
exponentially growing constraint violating modes. The problem of controlling the growth of the constraints for the conformal field equations has first been addressed by Florian Siebel in a diploma thesis \cite{florian_diploma}, and subsequently by H\"ubner and Siebel in \cite{Peter_Florian}. 
The key idea in this work is to develop
a $\lambda$--system \cite{BrFrHueRe99Ea}
for the conformal field equations in 1+1 dimensions (with toroidal ${\mycal I}$'s).
A  $\lambda$--system is an enlarged evolution system, where evolution
equations for the constraints are added in, consistently with symmetric
hyperbolicity. One then has a large parameter space of coefficient
functions available, in which to find choices such that the new system
has the constraint surface as an attractor.
The main conclusion of this work is that it was not possible to significantly
improve the fidelity of the numerical calculations. In those cases where
moderate improvements regarding the constraints could be achieved, the
deviation from the known exact solution would get larger.

\section{Results from 3D calculations}\label{sec:results}
%%%%%%%%%%%%%%%%%%%%%%%%%%%%%%%%%%%%%%

All the results presented in this section have been
performed with $121^3$ grids on 32 processors of the AEI's SGI origin 2000.
The outer boundary has been placed at a radius of $r = 1.15$ in these runs
(${\mycal I}^+$ is initially located at a radius $r = 1$).
% soem runs stopped after 8 hours in queue

\subsection{Minkowski data}\label{sec:minkdata}
%%%%%%%%%%%%%%%%%%%%%%%%%%%%%%%

We will first discuss some results for Minkowski spacetime, which in spite of
its simplicity provides some nontrivial numerical tests.
As has been first demonstrated by H\"ubner in \cite{PeterIV},
for weak data -- in particular Minkowski space -- it is possible via
the conformal approach to cover the whole domain of dependence of initial
data reaching out to ${\mycal I}^+$ with a finite number of time steps. Let us thus
first consider the gauges of Sec. \ref{sec:MinkTextbook},
where the compactified geometry is time-independent,
but a time-dependent conformal
factor $\Omega$ is responsible for contracting the cuts of ${\mycal I}^+$
to a point within finite coordinate time.

We have compared the gauges where the conformal spacetime
is Minkowski, (\ref{eq:OmegaMinkMink}), the Einstein static universe 
(\ref{eq:StandardGauge}), or the spacetime
given by  (\ref{eq:pseudostatic_spatiallyflat}). Essentially, the result is
that the Minkowski case yields the highest accuracy, the Einstein universe
case works in principle, and in the case (\ref{eq:pseudostatic_spatiallyflat})
the code crashes before reaching $i^+$.
\begin{figure}[htbp] % [b!] fig 1
\begin{center}
\includegraphics[width=2.35in]{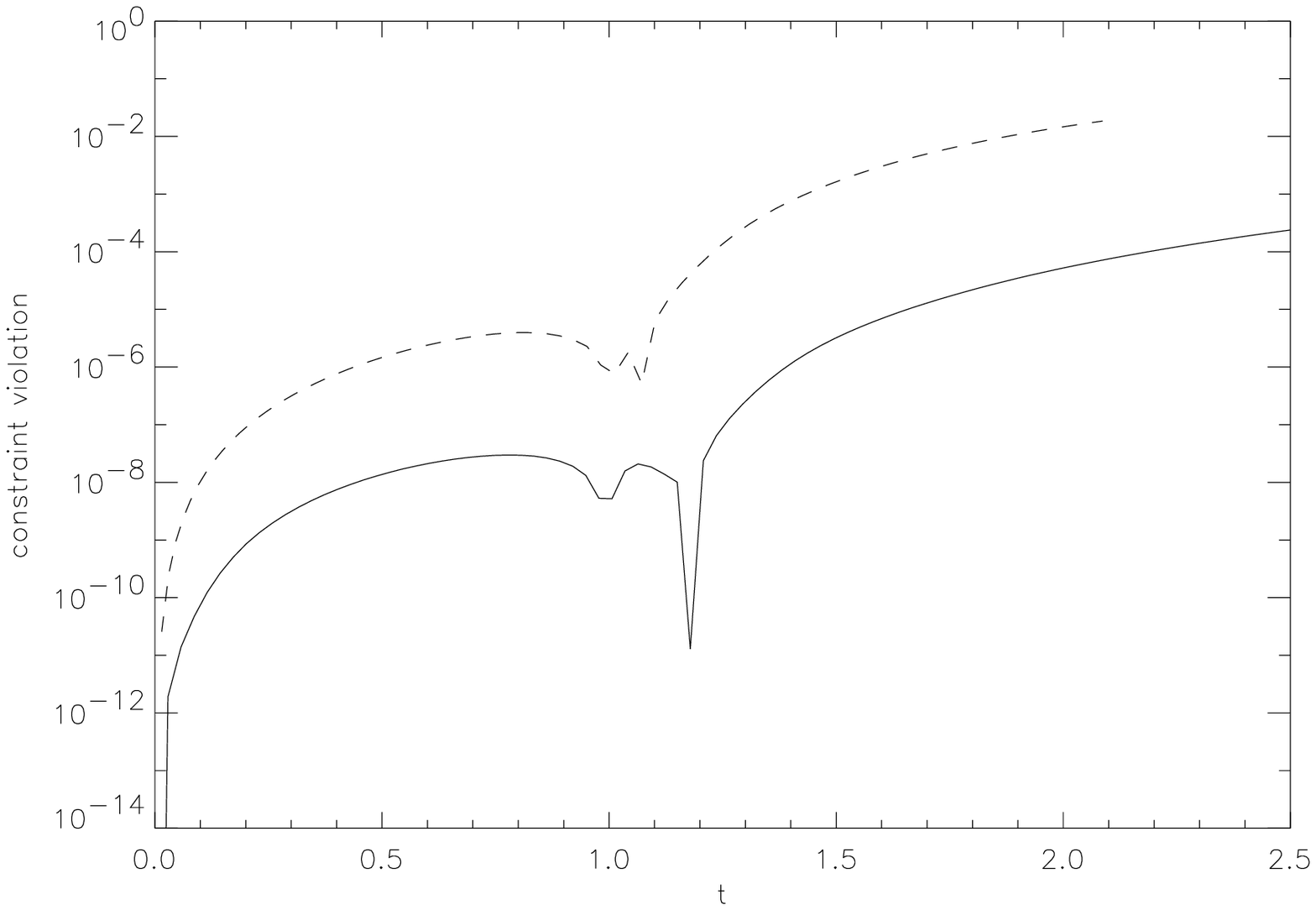}
\includegraphics[width=2.35in]{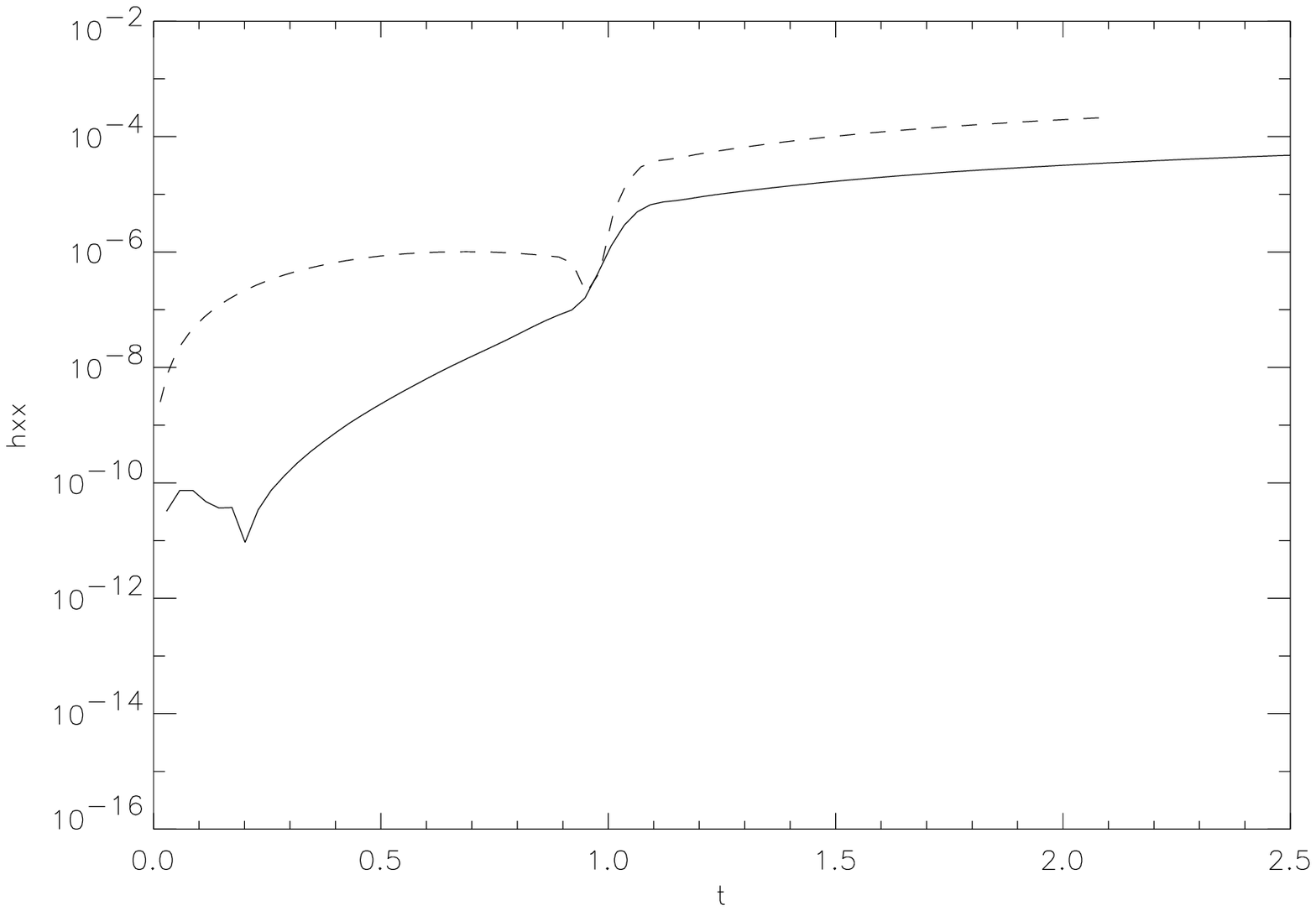}
\end{center}
%\epsfig{file=c73_StdMink_vsMinkMink.eps,height=2.35in,width=2.35in}
%\epsfig{file=hxx-hxx0_StdMink_vsMinkMink.eps,height=2.35in,width=2.35in}
%}
\caption{Comparing the Minkowski (solid line) and Einstein universe 
(dashed line) cases: left the value of the constraint
$\nabla_x \Omega = \Omega_x$ at the center is plotted
versus coordinate time, in the right image $h_{xx} - 1$ is plotted vs.
coordinate time (where $t$ is scaled such that $t(i^+)=1$)}
\label{fig:MinkMink_vs_StdMink}
\end{figure}
In Fig. \ref{fig:MinkMink_vs_StdMink} the Minkowski and Einstein universe
cases are compared by plotting $h_{xx} - 1$
and the value of the constraint $\nabla_x \Omega = \Omega_x$
at the center versus coordinate time
(where $t$ is scaled such that $t(i^+)=1$.
The Minkowski case -- denoted by the unbroken line --
clearly yields better accuracy, although the growth
of $h_{xx} - 1$ is faster, and approximately exponential during the
later stage of ``physical'' evolution. Note that the constraint grows
very fast in both cases.

\begin{figure}[htbp] % [b!] fig 1
\centerline{
\includegraphics[width=2.5in,height=2.35in]{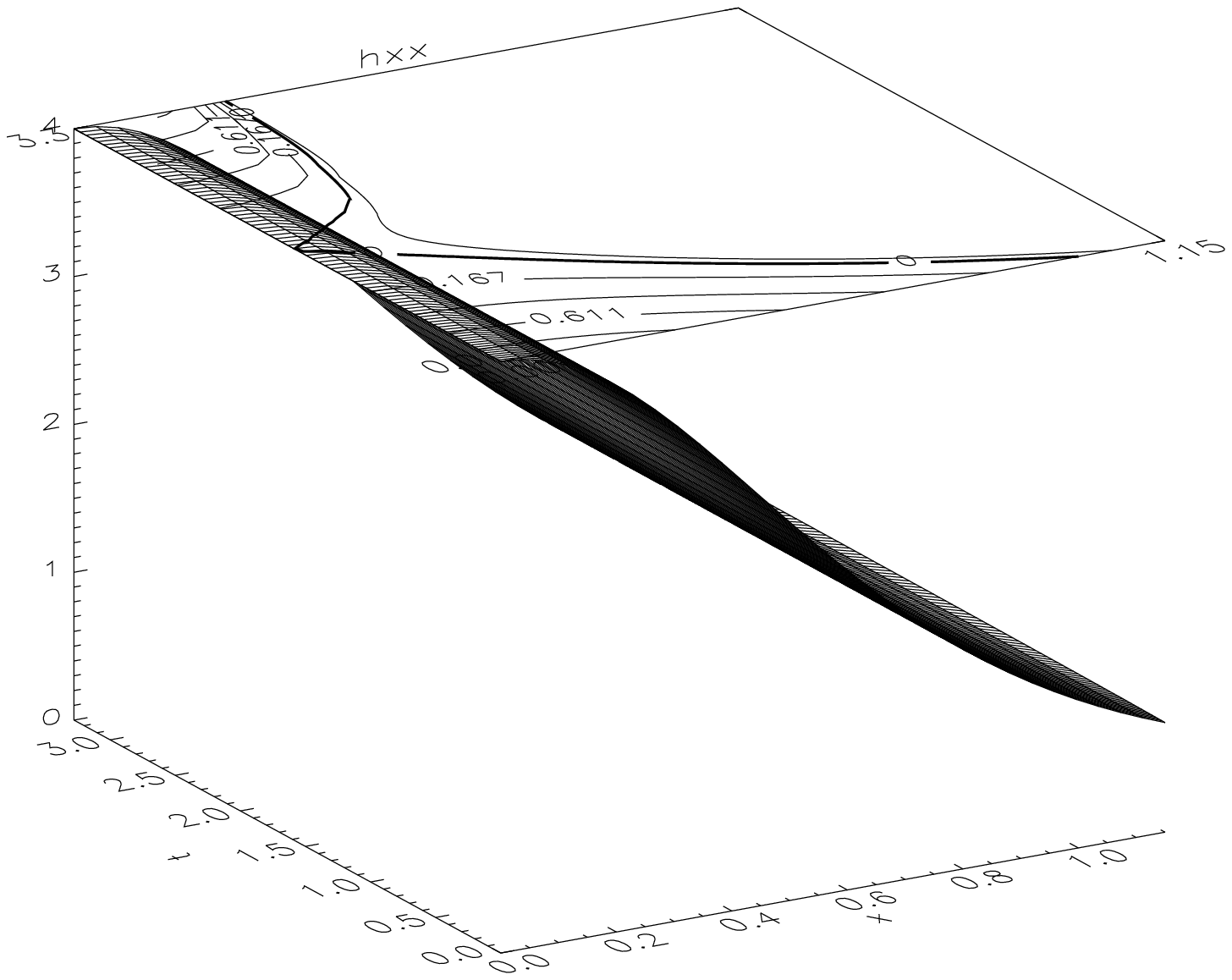}
\includegraphics[width=2.5in,height=2.35in]{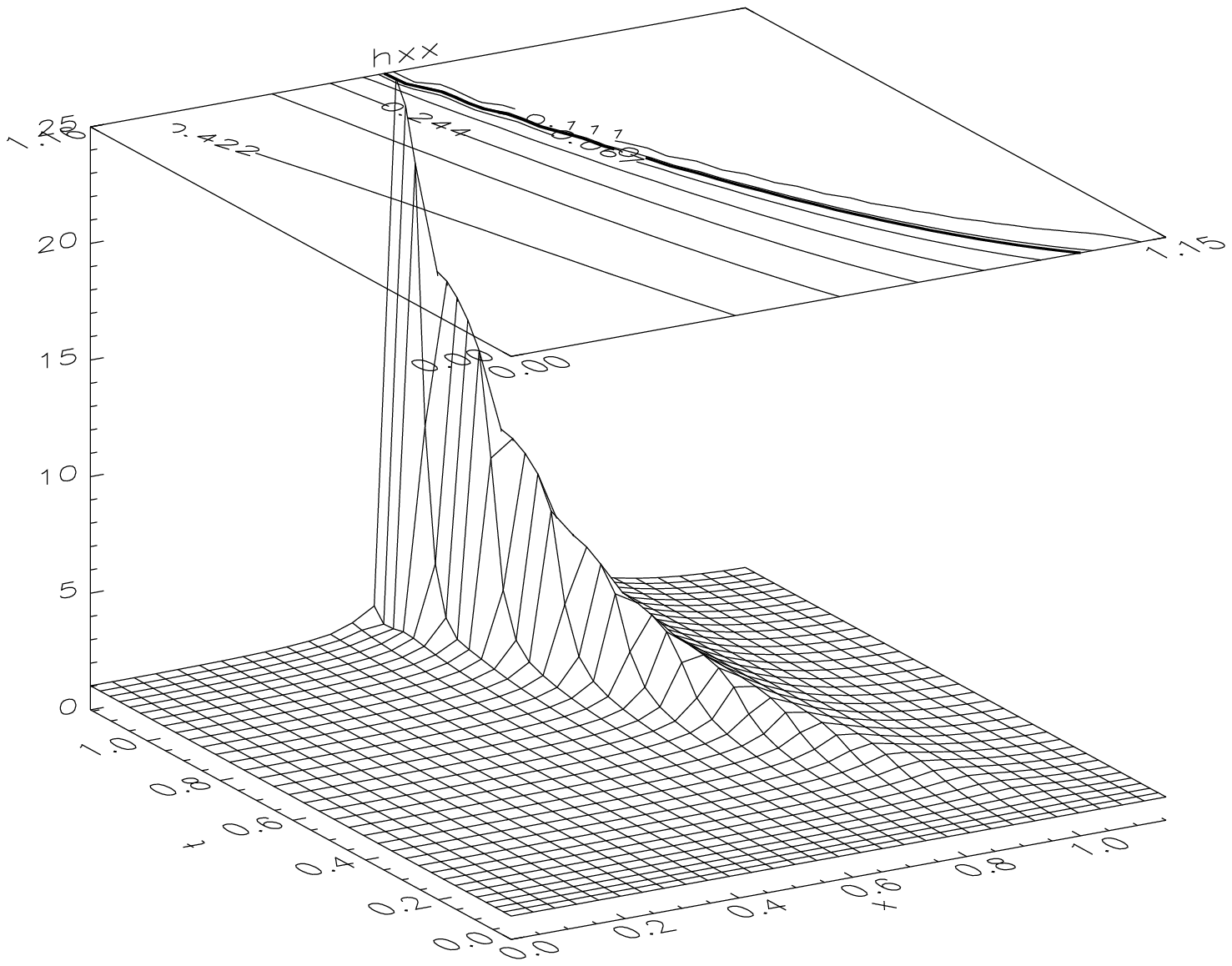}
} 
%\centerline{\epsfig{file=StandMink_var0.eps,         height=2.35in,width=2.35in}
%            \epsfig{file=BrillA0pseudoStaticVar0.eps,height=2.35in,width=2.35in}}
\caption{The value of the metric component $h_{xx}$ for $x\geq 0$ is plotted
versus coordinate time, the left image shows the Einstein universe case,
the right image shows case (\ref{eq:pseudostatic_spatiallyflat}), there
the maximum of $h_{xx}$ in the region where $\Omega > 0$ is approximately at
the value $5$}
\label{fig:StandMink_var0}
\end{figure}

\begin{figure}[htbp] % [b!] fig 1
\centerline{
\includegraphics[width=2.5in]{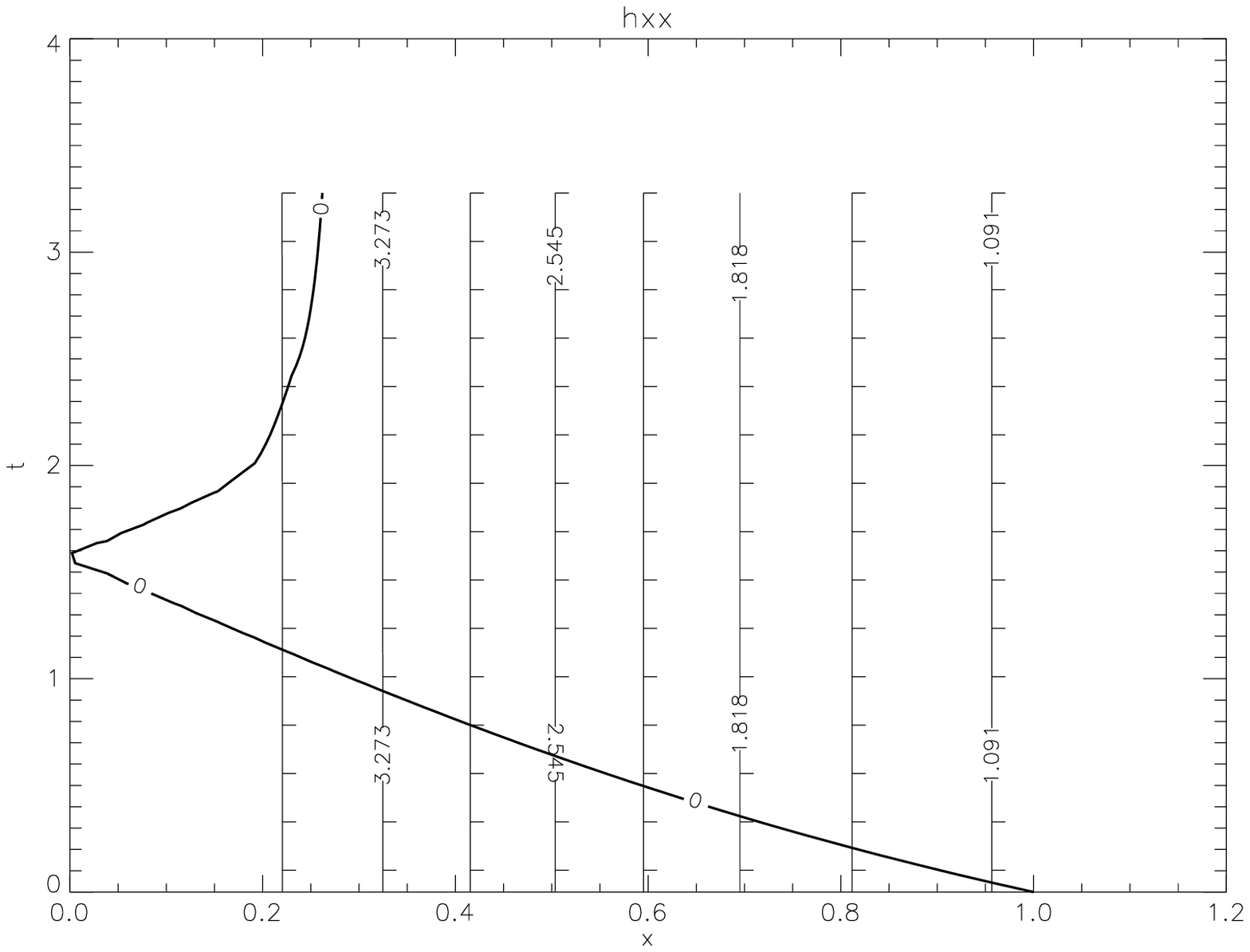}
\includegraphics[width=2.5in]{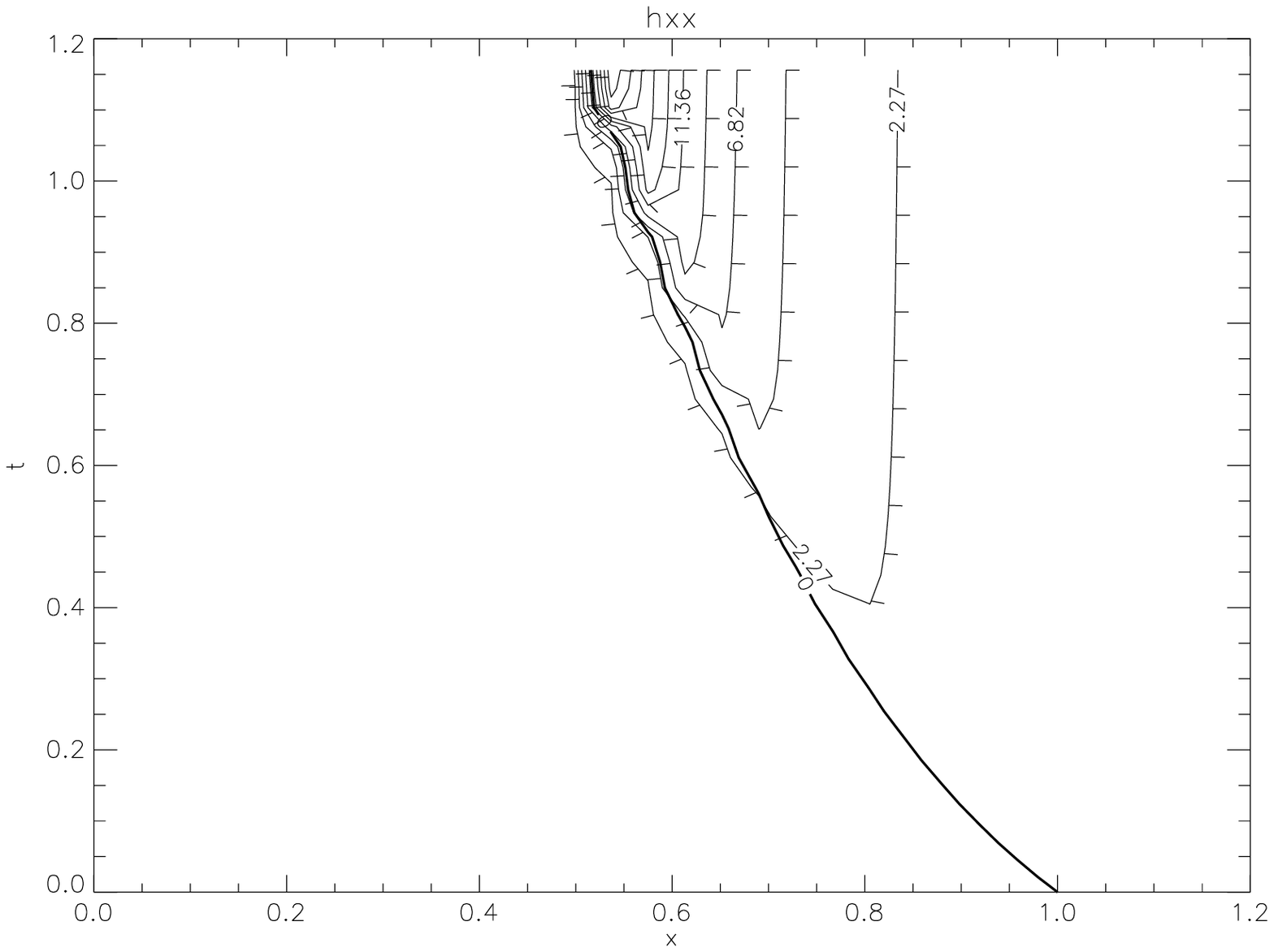}
}
%\centerline{
%\epsfig{file=StandMink_contourvar0.eps,height=2.35in,width=2.35in}
%\epsfig{file=BrillA0pseudoStaticContourVar0.eps,height=2.35in,width=2.35in}
%}
\caption{Contour lines of the metric component $h_{xx}$
for $x\geq 0$ are plotted
versus coordinate time, the left image shows the Einstein universe case,
the right image shows case (\ref{eq:pseudostatic_spatiallyflat}). The thicker
line marks $\Omega=0$, i.e. ${\mycal I}^+$}
\label{fig:Mink_contourVar0}
\end{figure}

\begin{figure}[htbp] % [b!] fig 1
\centerline{
\includegraphics[width=2.5in]{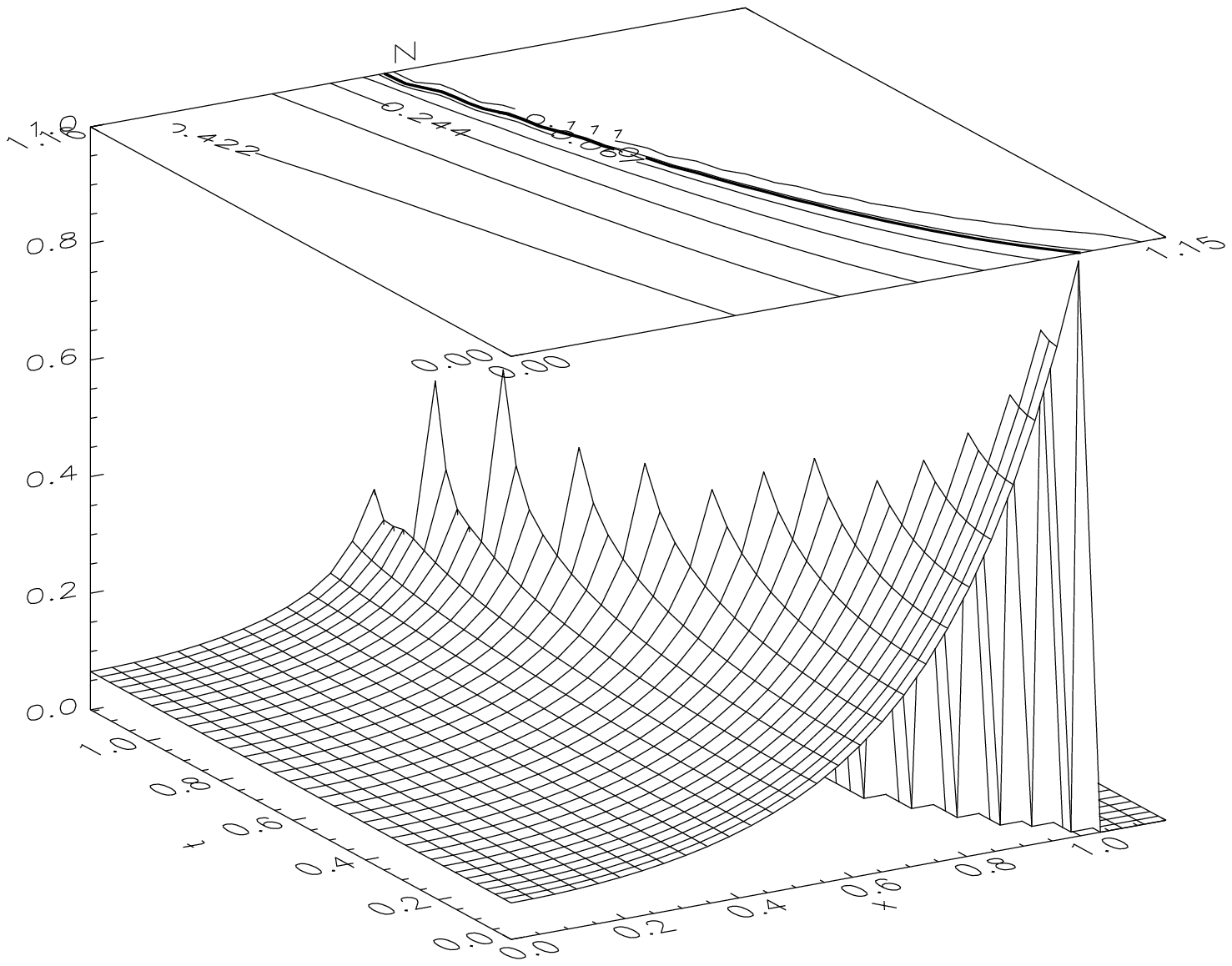}
\includegraphics[width=2.5in]{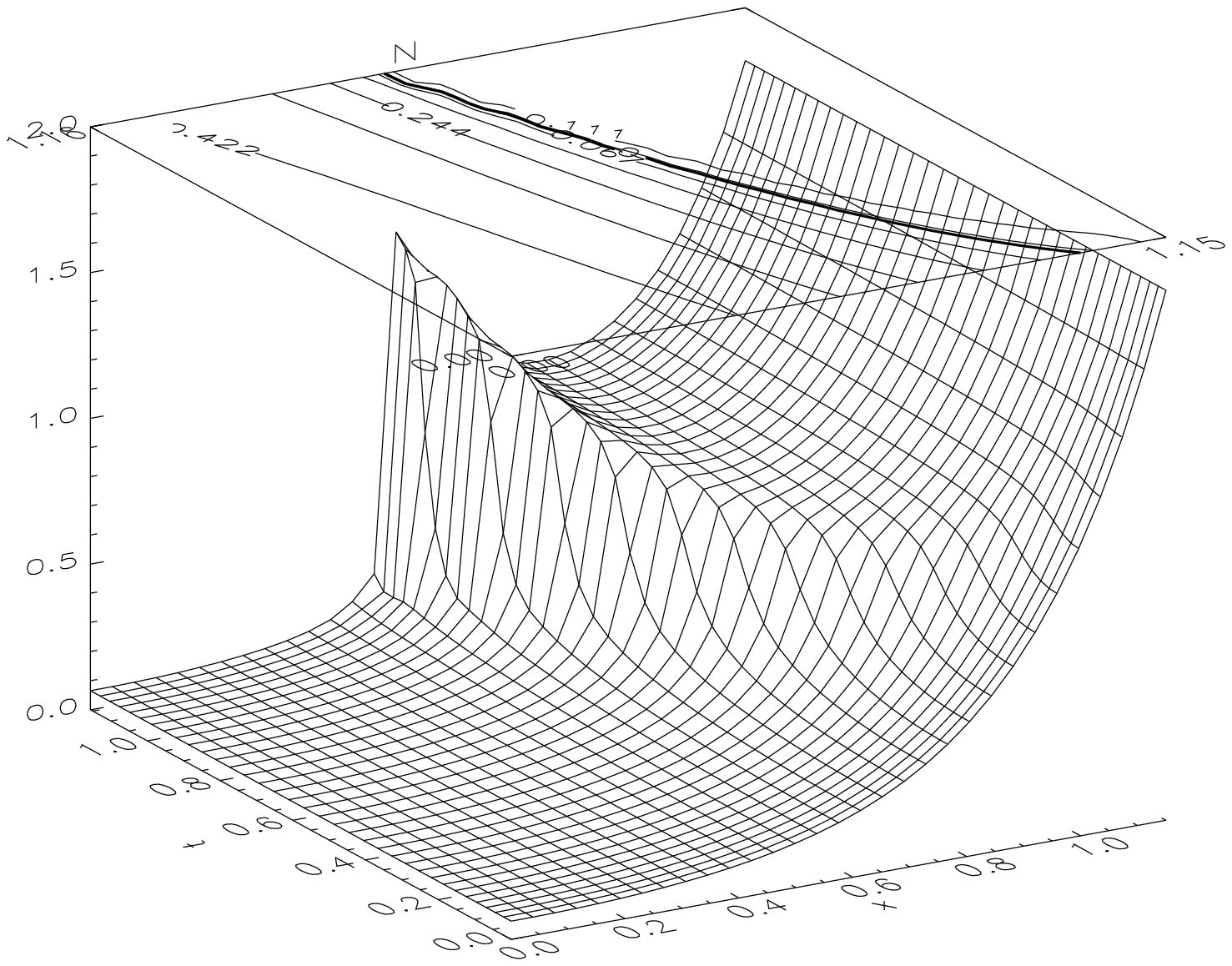}
}
%\centerline{
%\epsfig{file=BrillA0pseudoStaticCutVar57.eps,height=2.35in,width=2.35in}
%\epsfig{file=BrillA0pseudoStaticVar57.eps,height=2.35in,width=2.35in}}
\caption{The value of the lapse $N$ for $x\geq 0$ is plotted for the 
case (\ref{eq:pseudostatic_spatiallyflat})
versus coordinate time, the left image shows the points where $\Omega > 0$,
the right image shows all points}
\label{fig:mink:lapse}
\end{figure}

\begin{figure}[htbp] % [b!] fig 1
\centerline{
\includegraphics[width=2.5in]{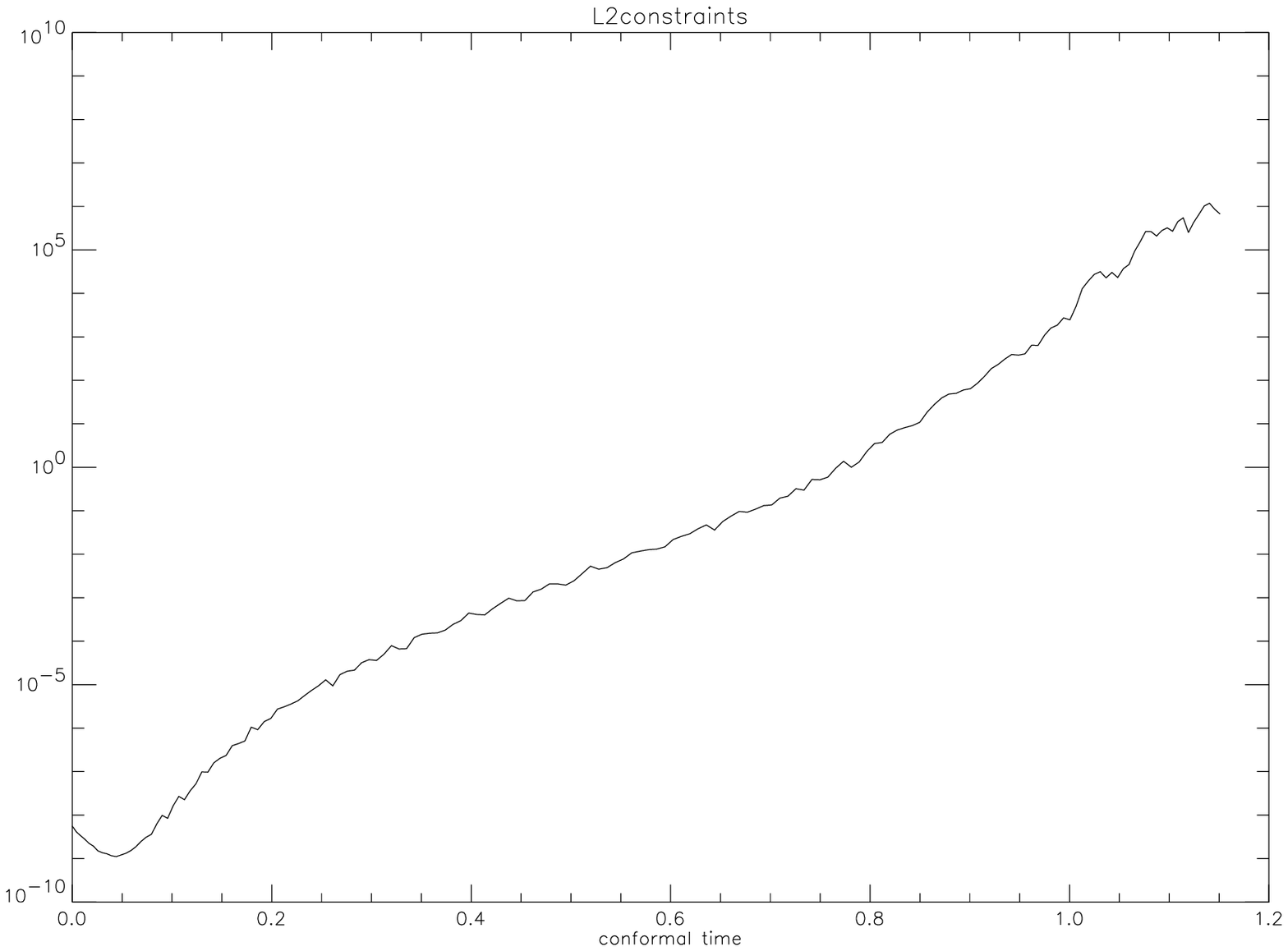}
\includegraphics[width=2.5in]{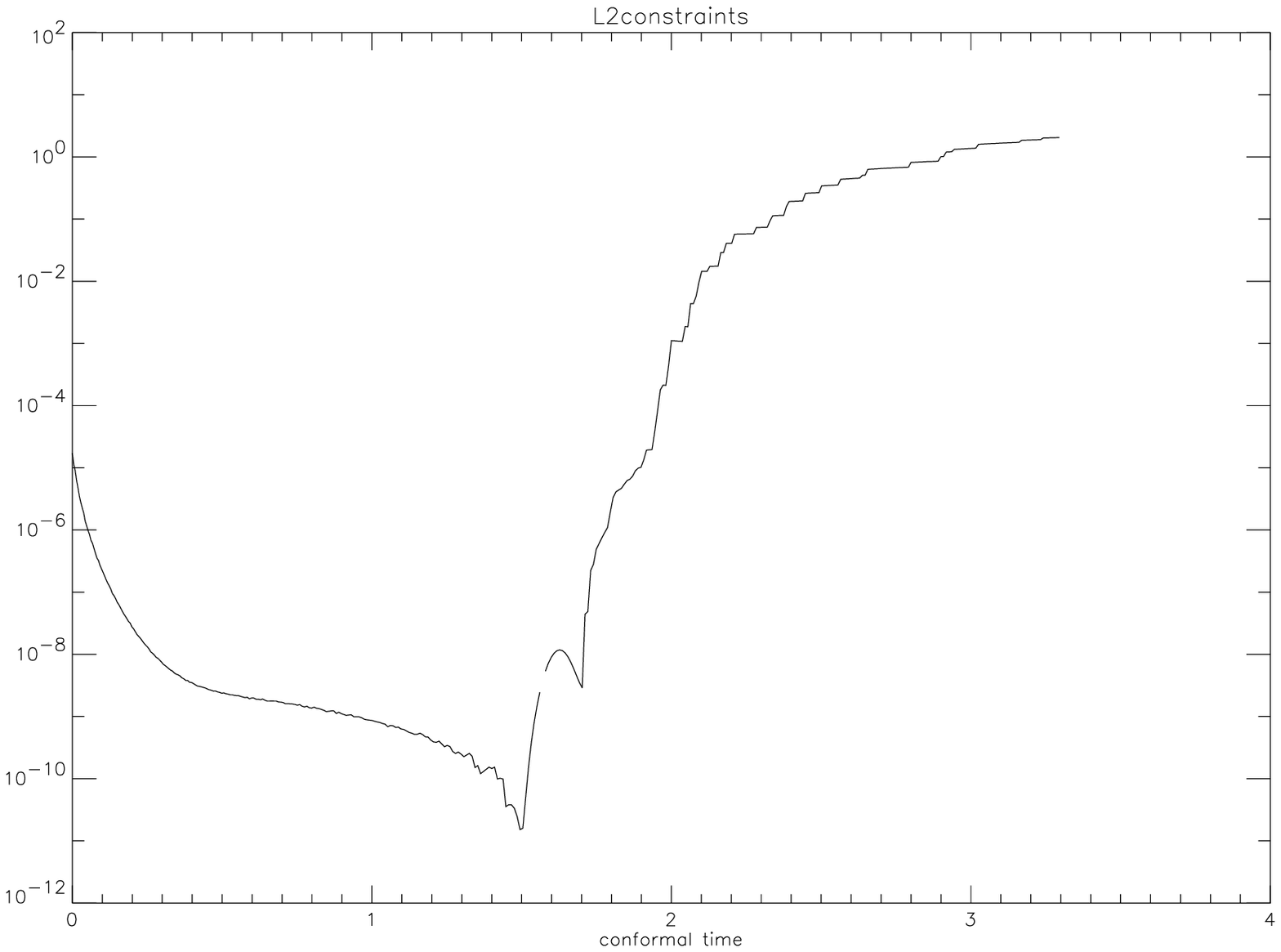}
}
%\centerline{
%\epsfig{file=BrillA0PseudStat-L2constraints-sum.eps,height=2.35in,width=2.35in}
%\epsfig{file=ConfBrillA0PseudStat-L2constraints-sum.eps,height=2.35in,width=2.35in}
%}
\caption{The sum over the $L^2$--norms (taken in the physical region) of all
the constraints is plotted versus coordinate time for the Einstein universe
(left) and case  (\ref{eq:pseudostatic_spatiallyflat})  (right)}
\label{fig:mink:L2constraints}
\end{figure}

Figures \ref{fig:StandMink_var0} -- \ref{fig:mink:L2constraints}
show a comparison of the less optimal Einstein universe case
with the case (\ref{eq:pseudostatic_spatiallyflat}) to illustrate
some of the problems one expects in the evolution of nontrivial spacetimes.
Fig. \ref{fig:StandMink_var0} shows the time evolution of $h_{xx}$ along
the positive $x$--axis versus coordinate time for the Einstein
universe case and for the case of  (\ref{eq:pseudostatic_spatiallyflat}).
Fig. \ref{fig:StandMink_var0} compares the corresponding contour lines.
While no deviation from staticity is visible for the Einstein universe case,
the other case shows a rapidly growing peak in $h_{xx}$ and the lapse
(shown in Fig. \ref{fig:mink:lapse})
(and thus of $\det h$), which is located in the transition zone outside of
${\mycal I}^+$.
Eventually this feature can not be resolved any more, and the code crashes.
In the Einstein static case the code was simply stopped by running out of time
in the queue. Fig. \ref{fig:mink:L2constraints} shows the sum over the
$L^2$--norms (taken in the physical region) of all the constraints versus time.
While in the Einstein static case the constraints show a rapid decrease
in the physical region, followed by a steep growth after passing through 
$i^+$, the case (\ref{eq:pseudostatic_spatiallyflat}) exhibits 
roughly exponential overall growth almost from the start.
%
%\begin{figure}[htbp] % [b!] fig 1
%\centerline{\epsfig{file=BrillA0pseudoStaticCutVar0.eps,height=2.35in,width=2.35in}
%\epsfig{file=BrillA0pseudoStaticVar0.eps,height=2.35in,width=2.35in}}
%\vspace{10pt}
%\caption{}
%\label{fig:BrillA0pseudoStaticVar0}
%\end{figure}
%

\begin{figure}[htbp] % [b!] fig 1
\centerline{
\includegraphics[width=2.5in]{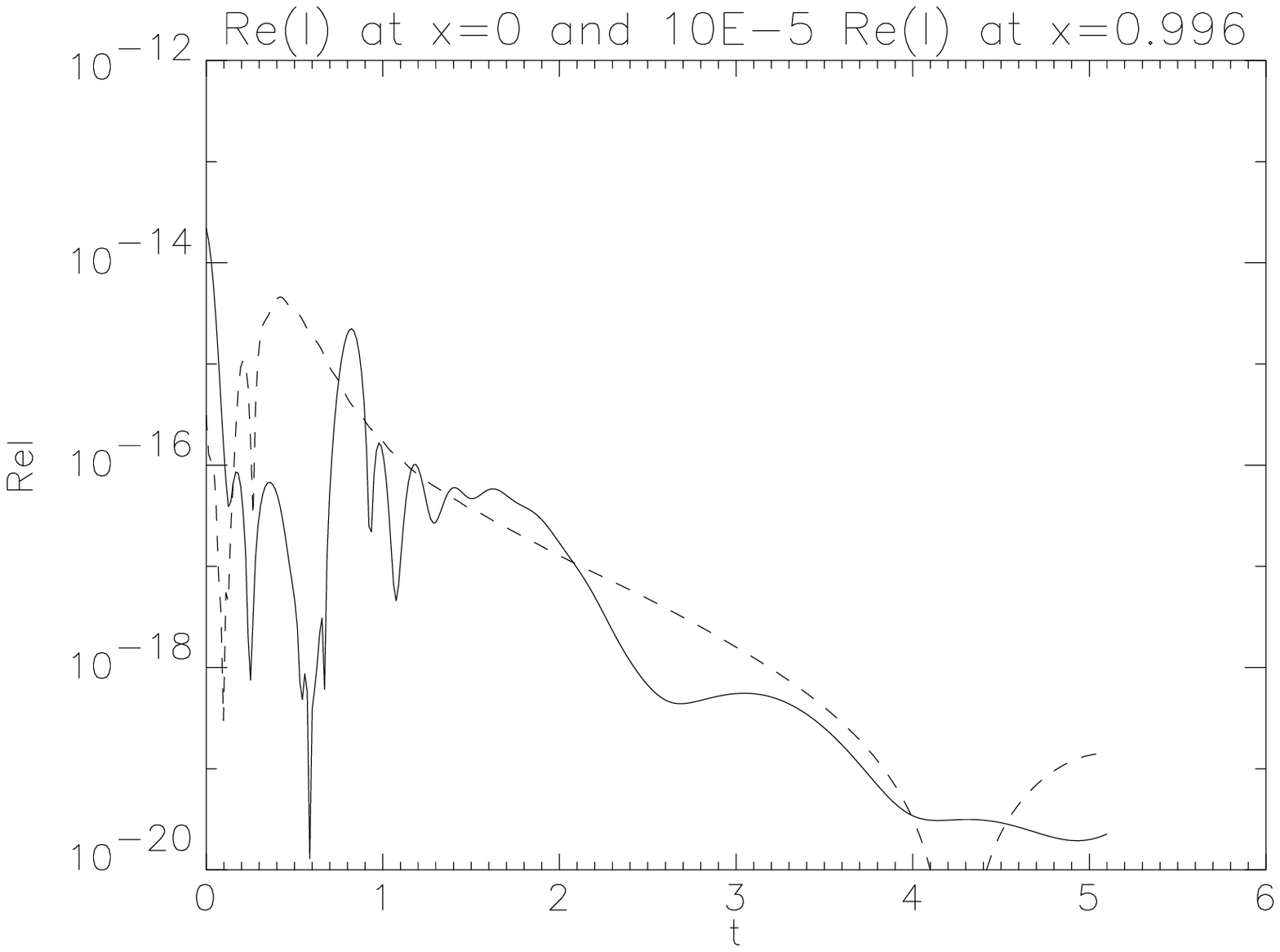}
\includegraphics[width=2.5in]{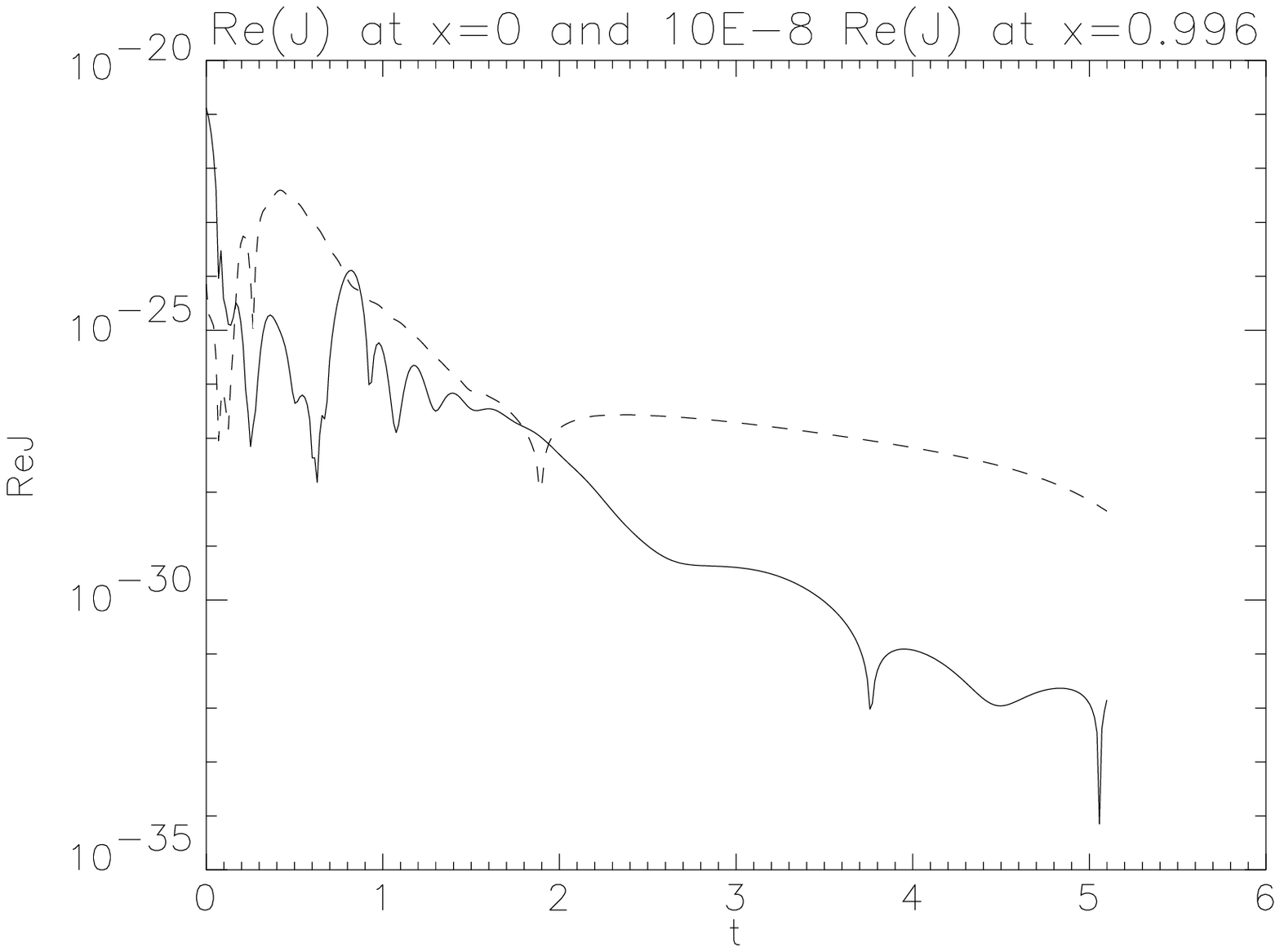}
}
%\centerline{
%\epsfig{file=ReI_line.eps,height=2.35in,width=2.35in}
%\epsfig{file=ReJ_line.eps,height=2.35in,width=2.35in}
%}
\caption{The real parts of the curvature invariants $I$ (left) and $J$ (right)
are plotted versus coordinate time  for the static gauge of  (\ref{eq:static_gauge}). The solid line is for the gridpoint
at the center of the grid, the dashed line for a grid point at
$x=0.996$, $y=z=0$, multiplied by a factor of $10^{-5}$ for $I$ and
$10^{-8}$ for $J$}
\label{fig:MinkStatic_ReI_ReJ}
\end{figure}

\begin{figure}[htbp] % [b!] fig 1
\centerline{
\includegraphics[width=2.5in]{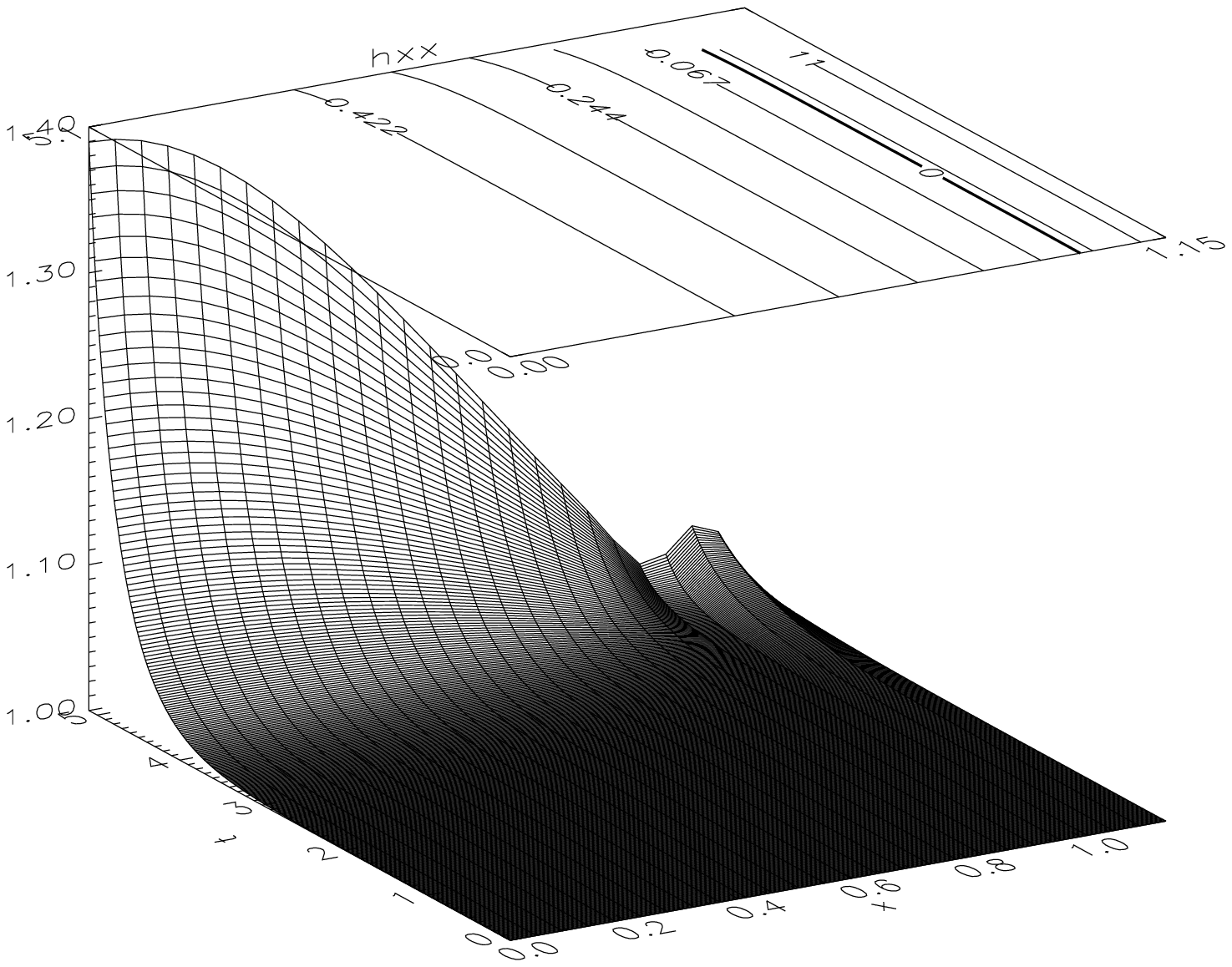}
\includegraphics[width=2.5in]{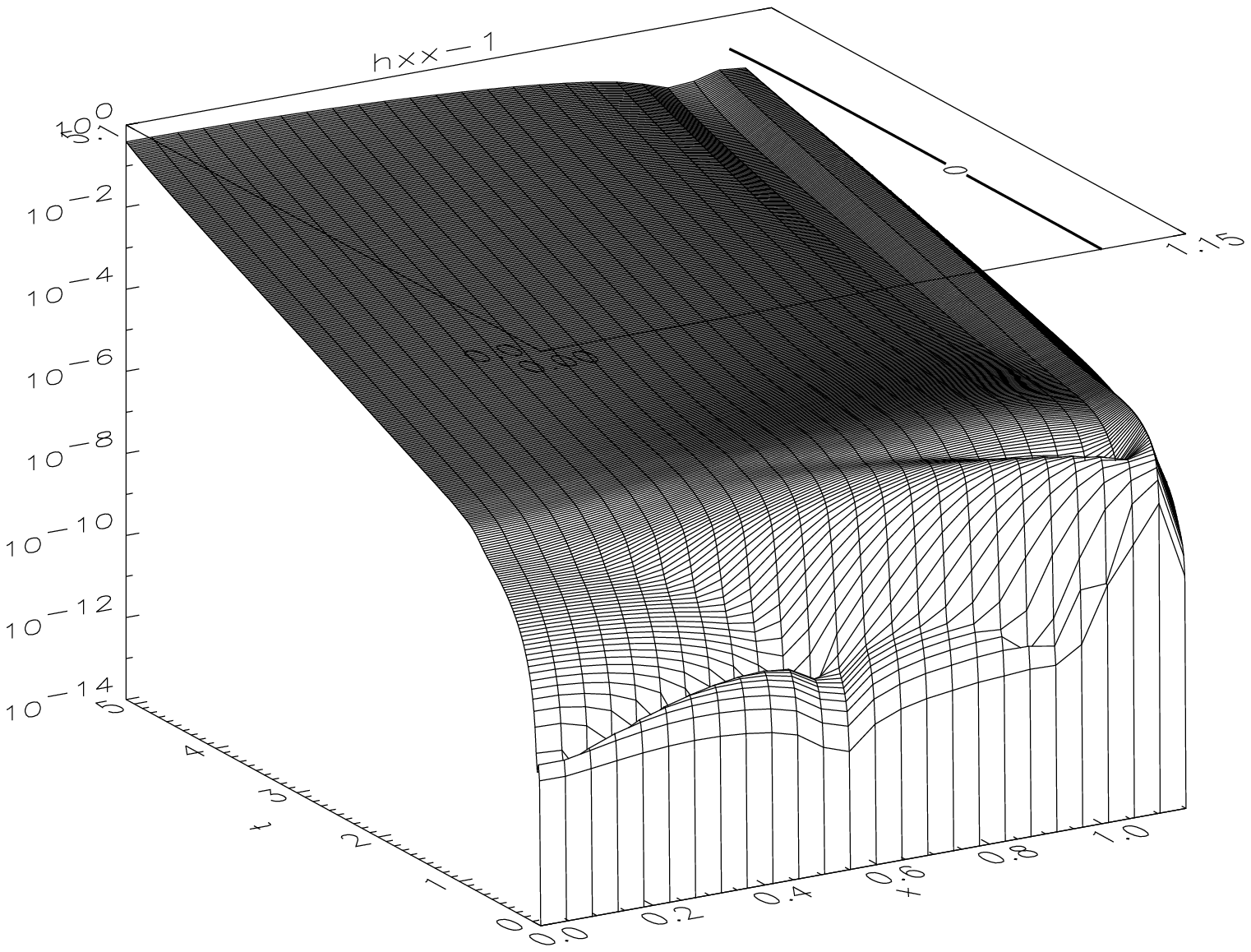}
}
%\centerline{
%\epsfig{file=hxx_StatMink.eps,height=2.35in,width=2.35in}
%\epsfig{file=hxx_StatMink_log.eps,height=2.35in,width=2.35in}
%}
\caption{The value of the metric component $h_{xx}$ for $x\geq 0$ is plotted
versus coordinate time with linear (left) and logarithmic (right) scaling
for the static gauge of  (\ref{eq:static_gauge}).
Approximately exponential growth is obvious, the largest amplitude of the
growth is in the center}
\label{fig:MinkStatic_hxx}
\end{figure}

\begin{figure}[htbp] % [b!] fig 1
\centerline{
\includegraphics[width=3.75in]
{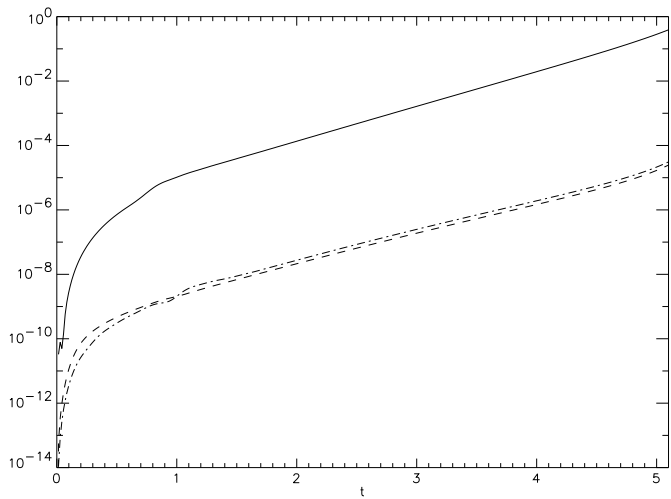}
}
%\centerline{
%\epsfig{file=hxx_cons73_cons0_StatMink_strichliert_is_c73.eps,height=2.35in,width=2.35in}}
\caption{The values of $h_{xx}$ (solid line) and the constraints  
$\nabla_x h_{xx}$  (dot--dashed) and $\nabla_x \Omega = \Omega_x$ (dashed)
are plotted versus coordinate time for the static gauge of
(\ref{eq:static_gauge})}
\label{fig:MinkStatic_exp_growth}
\end{figure}

Results for the completely static gauge given by  (\ref{eq:static_gauge})
are shown in Figs.
\ref{fig:MinkStatic_ReI_ReJ} -- \ref{fig:MinkStatic_exp_growth}.
This gauge poses a harder challenge than the previous ones, where
$i^+$ is reached in finite time. Now the goal is to maintain an indefinite
stable evolution. However, the evolution shows exponential growth,
illustrated in Figs. \ref{fig:MinkStatic_hxx} and \ref{fig:MinkStatic_exp_growth}  by the values of $h_{xx}$ and constraints
$\nabla_x h_{xx}$ and $\nabla_x \Omega = \Omega_x$.
Interestingly, however, the curvature invariants $I$ and $J$ are
{\em decreasing} during the evolution as shown in Fig.
\ref{fig:MinkStatic_ReI_ReJ}.
The exponential blowup crashes the code at $t\sim 5.1$,
this time seems to be roughly independent of resolution, size of time step,
amount of dissipation, location of the boundary and location
of the transition zone.
A possible explanation are exponentially growing constraint violating
modes on the continuum level.
%
%\begin{figure}[htbp] % [b!] fig 1
%\centerline{
%\includegraphics[width=2.5in]{ReI.eps}
%\includegraphics[width=2.5in]{ReJ.eps}
%}
%%\centerline{
%%\epsfig{file=ReI.eps,height=2.35in,width=2.35in}
%%\epsfig{file=ReJ.eps,height=2.35in,width=2.35in}
%%}
%\caption{The real parts of the curvature invariants $I$ (left) and $J$ (right)
%for $x\geq 0$ are plotted versus coordinate time  for the static gauge of  (\ref{eq:static_gauge}), superimposed are contour
%lines of the conformal factor $\Omega$}
%\label{fig:MinkStatic_ReI_ReJ}
%\end{figure}
%

\subsection{``Brill'' data}
%%%%%%%%%%%%%%%%%%%%

We use an axisymmetric Brill--wave type ansatz to look at initial
data that contain radiation and set
\begin{equation}\label{eq:Brill_ansatz}
\D s^2 = 
\omega^2\left(\E^{2 Q}(\D \varrho^2 +\D z^2) + \varrho^2 \D\varphi^2\right)\; ,
\end{equation}
where $\varrho^2 = x^2 + y^2$.
With $Q = \frac{1}{2} \ln (1 + A \bar\Omega^2 \varrho^2 f(\varrho^2))$,
%and the introduction of an auxiliary function
%$\bar Q = \varrho^{-2} \, e^{2 Q} - 1 =  \bar\Omega^2 f(\varrho^2)$, 
in  Cartesian coordinates the conformal three-metric becomes
\begin{eqnarray*}
    h_{B} & = &  \omega^2
      \left(\begin{array}{ccc}
    1 + A \, x^2 \, \bar\Omega^2 \, f  & A \, x y \, \bar\Omega^2 \, f  &  0 \\
        A \, x y \, \bar\Omega^2 \, f  & A \, y^2 \, \bar\Omega^2 \, f  &  0 \\
        0                & 0             &  1 + A \, \bar\Omega^2 \, f
      \end{array}\right)\; .
  \end{eqnarray*}
The axial symmetry makes it
easier to analyze the data and choose the gauges.
Here we set $\omega = f = 1$ and $A=1$.

\begin{figure}[htbp] % [b!] fig 1
\centerline{
\includegraphics[width=2.5in]{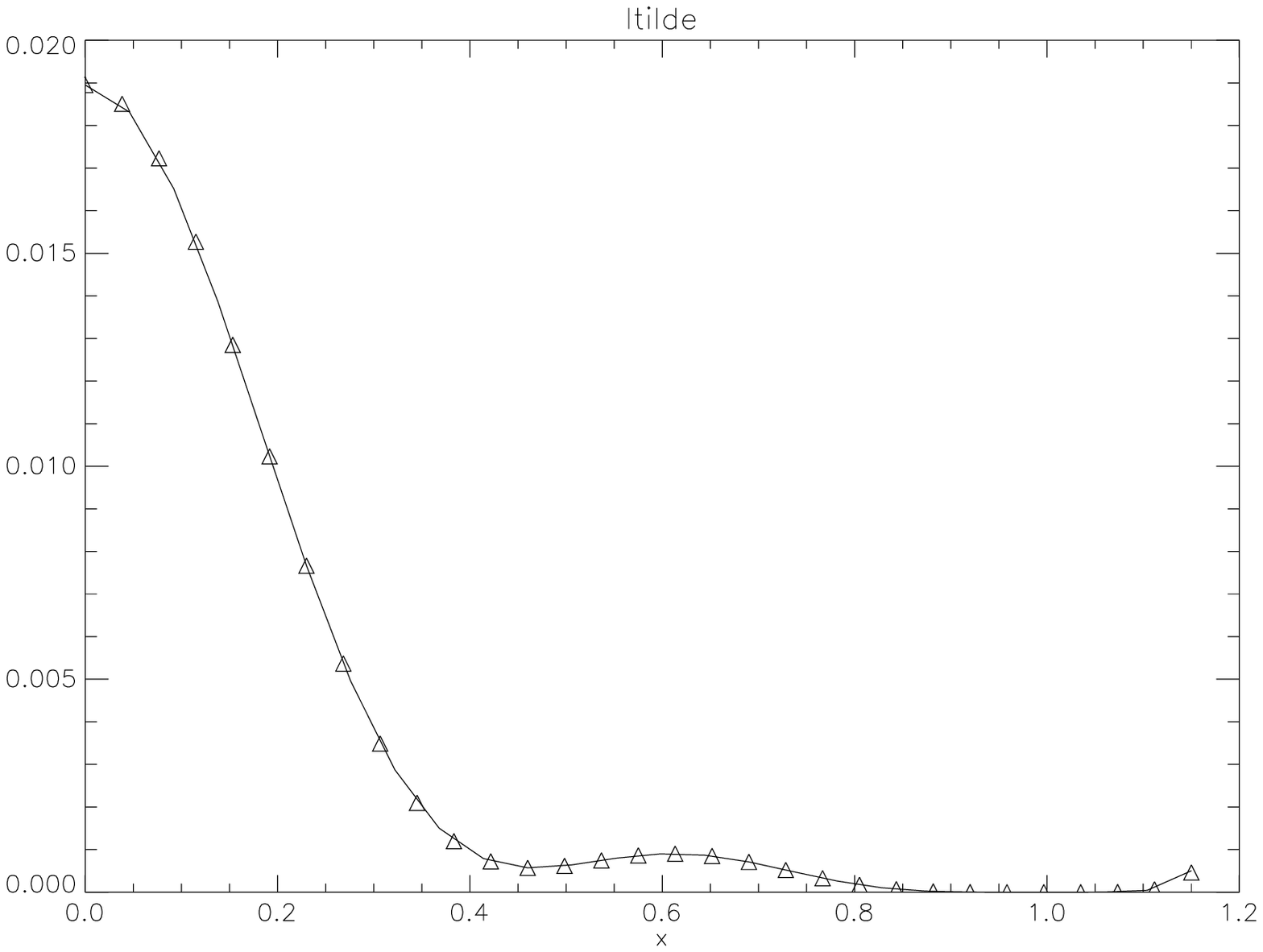}
\includegraphics[width=2.5in]{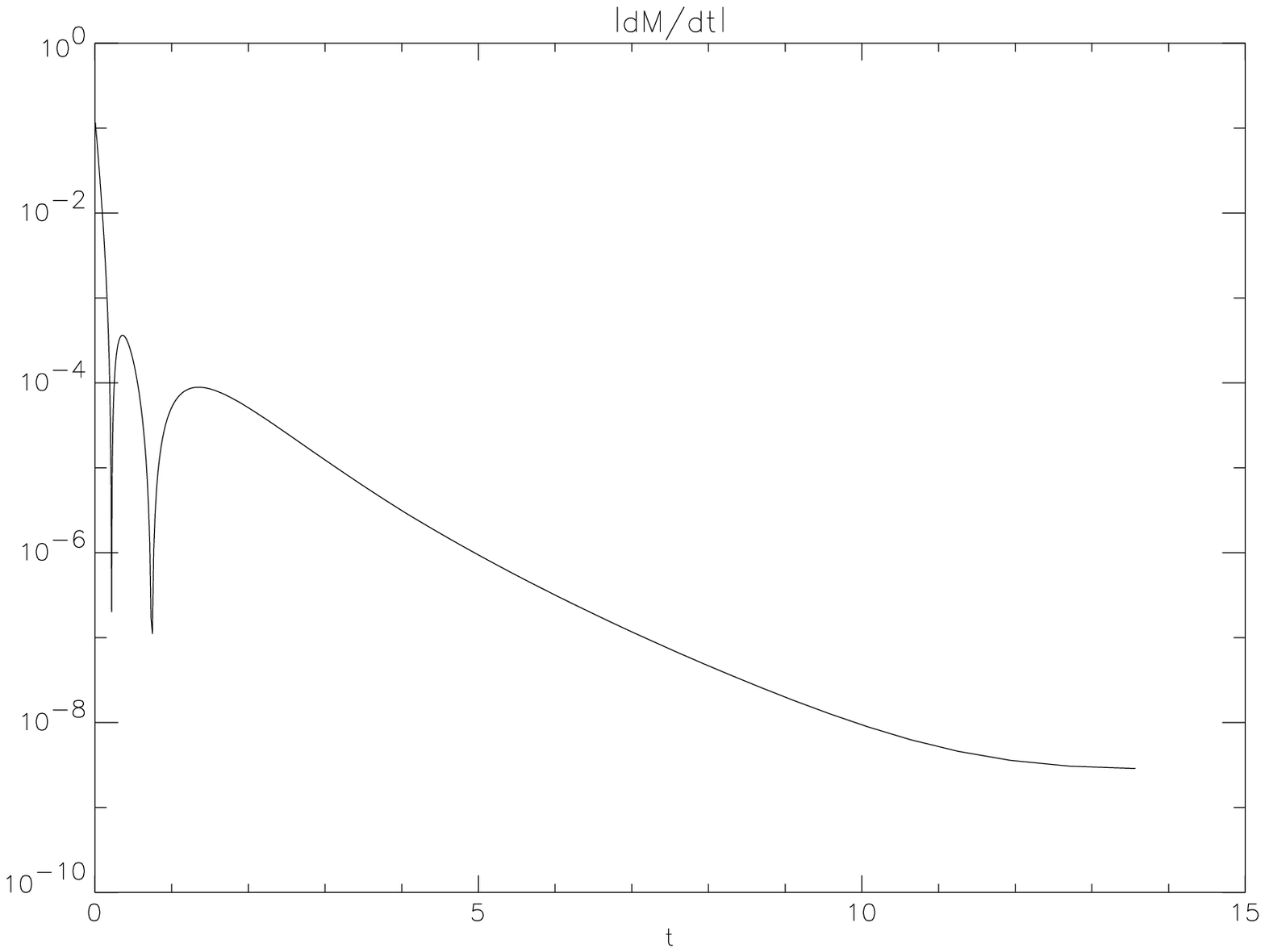}
}
%\centerline{
%\epsfig{file=Itilde_BrillvsConfBrill.eps,height=2.35in,width=2.35in}
%\epsfig{file=news_Brill_A1.eps,height=2.35in,width=2.35in}
%}
\caption{The left image shows the real part of the physical curvature
invariant $\tilde I = \Omega^6 \, I$, computed as a perturbation
of the Einstein universe (line) and case
(\ref{eq:pseudostatic_spatiallyflat})(triangles)
for a ``Brill wave'' with $A=1$.
The right image shows the corresponding mass loss function $\dot M_B$,
computed as as a perturbation of the Einstein static case ($R_g=6$)}
\label{fig:A1_Itilde_and_news}
\end{figure}

Fig. \ref{fig:A1_Itilde_and_news} shows
the real part of the physical curvature
invariant $\tilde I = \Omega^6 \, I$ and the mass loss $\dot M_B$.
The curvature invariant  $\tilde I$ is computed both as a perturbation
of the Einstein universe and case
(\ref{eq:pseudostatic_spatiallyflat})(triangles)
for a ``Brill wave'' with $A=1$, to demonstrate the the physical
initial data are indeed identical.
The mass loss $\dot M_B$ is computed as
a perturbation of the Einstein static case ($R_g=6$) and plotted
in a logarithmic scale. Note that the
falloff levels off at late times to a constant value due to
numerical error. Note also that oscillations, like present here, are absent
from the initial data corresponding to  (\ref{eq:standard-h}) as shown
in Fig. 5 of \cite{my_tuebingen_LNP}.

\section{Conclusions and Outlook}\label{sec:discussion}
%%%%%%%%%%%%%%%%%%%%%%%%%%%%%%%%%%%%%%%%%

Bringing the conformal approach to numerical relativity to full fruition
such that it can be used as a tool to
explore new physics -- in particular in black hole spacetimes -- will be a long
term effort.
In order to contemplate the scope of this project, let us give a drastically
oversimplified definition of the art of numerical relativity as a
procedural recipe:
\begin{enumerate}
\item Find a well posed formulation of the initial(-boundary) value and 
initial data (constraint) problems for general relativity
(optimally, well-posedness should be a theorem but
good numerical evidence may be considered sufficient).
\item Without destroying well-posedness, modify your equations and
choose your gauges, such that your problem actually becomes well-conditioned
\footnote{Ill conditioned problems are those where a result depends very
strongly on input, i.e. on initial data, 
see e.g. Sec. 1.6 and 6.1 of \cite{Atkinson}.}.
\item Construct a solid numerical implementation, flexible enough to
handle experiments as required by science and by finding solutions
to the problems associated with point two.
\item Discover (new) results in physics.
\item Explain what you achieved (and how) to fellow numerical 
relativists and others, such as mathematical
relativists, astrophysicists, cosmologists, or mathematicians. 
\end{enumerate}
Even without considering the last point (which the present
article humbly tries to serve), numerical relativity
is a challenging enterprise.

The conformal approach complies with point one in the optimal sense:
the equations are regular in the whole spacetime, including
the asymptotic region, there are no ambiguities associated with ad-hoc
cutoffs at finite distance, and the evolution equations
are symmetric hyperbolic, which
guarantees well-posedness of the initial value problem and allows
to obtain well-posed initial-value-boundary problems.

Point two however already poses a significant challenge:
Well-defined is not well-conditioned,
well-posed problems may still be hopelessly ill-conditioned for numerical
simulation. A simple example is provided by any chaotic dynamical system
(in the sense of ordinary differential equations). 
When it comes to solving the Einstein equations, the gauge freedom of the 
theory results in having more equations (constraints and evolution equations)
than variables, and more variables than physical degrees of freedom. This
redundancy can easily lead to spurious approximate solutions.
Different ways to write the equations are only equivalent with regard to exact
solutions, but approximations will tend to exhibit constraint violating or
gauge modes that may grow very fast (e.g. exponentially). This is perfectly
consistent with well-posedness but not acceptable numerically.
Even without triggering instabilities, the choice of a bad gauge
is likely to create features in the solution which are 
in practice impossible to resolve.
The ``good news'' is that many of the problems encountered with
the conformal field equations have counterparts in traditional approaches
to numerical relativity. The way toward solving these problems usually
takes the form of gaining insight from simplifications and
analytical studies, which then have to be tested in full numerical
simulations. This requires a flexible code that is geared toward
performing the necessary experiments, which leads to point three --
another hard task for classical relativists, because
it requires an engineering attitude many relativists are not familiar with.
%Point four seems to be what physicists are particularly trained for
%(experimental physicists are usually also trained for point three),
The gauge freedom of general relativity and absence of
a natural background creates an additional twist when it comes to point four,
which leads to numerous technical and conceptual subtleties.

What is the roadmap for the future? In order to
comply with points two and three of the above recipe, preliminary work is 
carried out toward a new 3D code that will be flexible enough to carry out
a range of numerical experiments in order to come up with
well-conditioned algorithms for the conformal field equations.
One major issue in the improvement of algorithms is to implement
a better boundary condition, that does not require a transition zone,
allows the boundary to be closer to ${\mycal I}^+$ and minimize
constraint violations generated at the boundary or outside ${\mycal I}$.
Here an essential problem is that ${\mycal I}^+$ has spherical cuts, and
algorithms based on Cartesian grids are probably not optimal.
Certainly, a lot of energy will have to be devoted to the question of
finding appropriate gauge conditions. Particularly hard seems to be
the question of how to choose the Ricci scalar $R$ of the unphysical
spacetime. Since $R$ steers the conformal factor implicitly through
nonlinear PDEs, it seems very hard to influence the
conformal factor in any desired way.

An important role in improving the analytical understanding and in setting
up numerical experiments will be played by the utilization of
simplifications. Particularly important are spacetime symmetries
and perturbative studies.
Minkowski and Kruskal spacetimes provide particularly important cases
to be studied in this context.
An alternative route to simplification, which has been very successful
in numerical relativity, is perturbative analysis, e.g. with Minkowski
or Schwarzschild backgrounds. In the context of compactification
this has been carried out numerically
with characteristic codes in \cite{pertI,pertII}
(using appropriate variables in the Teukolsky
equations, the perturbation equations are made regular at ${\mycal I}^+$),
some of the problems
that showed up there are likely to be relevant also for the conformal approach.

The theory of general relativity is known as a never drying out source
for subtle questions in physics and mathematics.
Numerical relativity is hoped to help answer some important questions
-- but  at the same time poses many new ones.
Without a thorough understanding of how to obtain approximate solutions,
our insight into the theory seems incomplete. For isolated systems, 
the mastering of compactification techniques promises reliability and
precision. The next years are expected to
see some significant progress in this direction.

\section*{Acknowledgments}
The author thanks R. Beig, J. Frauendiener, H. Friedrich, B. Schmidt,
and M. Weaver for helpful discussions,
C. Lechner and J. Valiente Kroon for a careful reading of the manuscript,
and P. H\"ubner and M. Weaver for letting him use
their codes, explaining their results and providing general support in order
to take over this project.

%INDEX%%%%%%%%%%%%%%%%%%%%%%%%%%%%%%%%%%%%%%%%%%%%%%%%%%%%%%%%%%%%%%%
% Please check with the editor of your book whether he plans to
% include a "mutual" subject index - if so, please code your entries
% in the standard syntax. For your own purposes you may print your
% "personal" index by using the following commands:
%
%\clearpage
%\addcontentsline{toc}{section}{Index}
%\flushbottom
%\printindex
%%%%%%%%%%%%%%%%%%%%%%%%%%%%%%%%%%%%%%%%%%%%%%%%%%%%%%%%%%%%%%%%%%%%%

\end{document}